\documentclass[twocolumn,secnumarabic,amssymb, nobibnotes, aps, 10pt, floatfix, showpacs, prb]{revtex4-1}
\usepackage{graphicx}
\usepackage{dcolumn}
\usepackage{bm}
\usepackage{amsmath} 
\usepackage{subfigure}
\usepackage{url}

\graphicspath{{Figuren/},{Figuren/eta=0/},{Figuren/stefantda/},{Figuren/bigproblem/},{Figuren/xi=0.4picketfence/},{Figuren/readgreen/},{Figuren/gapecht/},{../../../RG/picketfenceeta1excitedphole/},{../../../RG/etachangepicket/}}

\newcommand{\ket}[1]{\vert #1 \rangle}

\newcommand{\da}[1]{#1^{\dagger}}
\newcommand{\com}[2]{\left[ #1, #2 \right]}

\begin{document}

\title{Exact solution of the $p_x + i p_y $ pairing Hamiltonian by deforming the pairing algebra}
\author{Mario Van Raemdonck}
\email{mariovr5@hotmail.com}
\affiliation{Ghent Quantum Chemistry Group, Ghent University, Krijgslaan 281 (S3), 9000 Gent, Belgium}
\affiliation{Center for Molecular Modeling, Ghent University, Technologiepark 903, 9052 Zwijnaarde, Belgium}
\author{Stijn De Baerdemacker}
\affiliation{Center for Molecular Modeling, Ghent University, Technologiepark 903, 9052 Zwijnaarde, Belgium}
\author{Dimitri Van Neck}
\affiliation{Center for Molecular Modeling, Ghent University, Technologiepark 903, 9052 Zwijnaarde, Belgium}

\date{\today}
\begin{abstract}
Recently, interest has increased in the hyperbolic family of integrable Richardson-Gaudin (RG) models. It 
was pointed out that a particular linear combination of the integrals of motion of the hyperbolic RG model
leads to a Hamiltonian that describes $p$-wave pairing in a two dimensional system. 
Such an interaction is found to be present in fermionic superfluids ($^3\mbox{He}$), ultra-cold atomic gases and $p$-wave superconductivity.
Furthermore the phase diagram is intriguing, with the presence of the Moore-Read and Read-Green lines. At the Read-Green line a rare third-order quantum phase transition occurs.  
The present paper makes a connection between collective bosonic states and 
the exact solutions of the $p_x + i p_y$ pairing Hamiltonian. This makes it possible to investigate the effects of the Pauli principle on the energy spectrum,
by gradually reintroducing the Pauli principle. It also introduces an efficient and stable numerical method to probe
all the eigenstates of this class of Hamiltonians.
We extend the phase diagram to repulsive
interactions, an area that was not previously explored due to the lack of a proper mean-field solution in this region.
We found a connection between the point in the phase diagram where the ground state connects to 
the bosonic state with the highest collectivity, and the Moore-Read line where all the Richardson-Gaudin (RG) variables collapse to zero. 
In contrast with the reduced BCS case, the overlap between the ground state and the highest collective state at the Moore-Read line is not the largest. In fact it
shows a minimum when most other bosonic states show a maximum of the overlap. 
We found remnants of the Read-Green line for finite systems, by investigating the total spectrum.
A symmetry was found between the Hamiltonian with and without single-particle part. 
When the interaction is repulsive we found 4 different classes of trajectories of the RG variables. 
\end{abstract}

\pacs{02.30.Ik , 74.20.Fg , 74.20.Rp , 71.10.Hf}
\maketitle

\section{Introduction}
Pairing plays an important role in the description of many phenomena as diverse as superconductivity in condensed matter systems \cite{bardeen:1957}, neutron stars \cite{dean:2003}, and
the interaction of nucleons in atomic nuclei \cite{dean:2003}. Probably the most notorious Hamiltonian that describes paired fermions is 
the reduced BCS Hamiltonian \cite{bardeen:1957}, which has an exact Bethe ansatz solution obtained by Richardson in 1963 \cite{richardson:1963}. 
The Richardson-Gaudin (RG) model
belongs to a more general class of integrable Hamiltonians, \cite{dukelsky:2004,ortiz:2005} which can be categorized into three families: the rational (or XXX), hyperbolic (or XXZ) and elliptic (or XYZ) RG models.
The reduced BCS Hamiltonian is part of the rational family. The rational model has attracted more interest during the last decade because
it describes pairing correlations in finite-size (mesoscopic) metallic nanograins \cite{vondelft:2001}. 
This has lead to applications in superconductivity\cite{vondelft:2001}, quantum optics\cite{dukelsky:2004}, cold-atomic physics\cite{dukelsky:2004}, quantum dots\cite{dukelsky:2004}, etc. The other two families
remained obscure until recently applications for the hyperbolic model were found in the context of $p$-wave pairing in ultra-cold Fermi gases \cite{nishida:2009,zhang:2008},
exotic superconductors such as $\mbox{Sr}_2\mbox{Ru}\mbox{O}_4$ \cite{xia:2006} and in the context of pairing in heavy nuclei. \cite{dukelsky:2011}.
The long standing importance of $p$-wave pairing in the $^3\mbox{He}$ superfluid state \cite{gurarie:2007} should also be kept in mind. \par 
Two-dimensional $p$-wave pairing can be described by means of the $p+ip$ Hamiltonian
\begin{equation}
H=\sum_{\mathbf{k}} \frac{\mathbf{k}^2}{2m}c^\dag_{\mathbf{k}}c_{\mathbf{k}} -\frac{G}{4m} \sum_{\mathbf{k}\mathbf{k}^\prime} \mathbf{k}\cdot\mathbf{k}^\prime c^\dag_{\mathbf{k}} c^\dag_{-\mathbf{k}} c_{-\mathbf{k}^\prime} c_{\mathbf{k}^\prime},
\end{equation}
with $\mathbf{k}\cdot\mathbf{k}^\prime=k_x k_x^\prime + k_y k_y^\prime$.  Its "chiral" variant
\begin{equation}
H=\sum_{\mathbf{k}} \frac{\mathbf{k}^2}{2m}c^\dag_{\mathbf{k}}c_{\mathbf{k}} -\frac{G}{4m} \sum_{\mathbf{k}\mathbf{k}^\prime} (k_x-ik_y)(k_x^\prime+ik_y^\prime) c^\dag_{\mathbf{k}} c^\dag_{-\mathbf{k}} c_{-\mathbf{k}^\prime} c_{\mathbf{k}^\prime},
\label{bham}
\end{equation}
also referred to as the $p_x+ip_y$ Hamiltonian, essentially captures the same physics \cite{foster:2013}, and is derivable as a linear combination of the
integrals of motion of the hyperbolic RG model \cite{dukelsky:2001} (see section(\ref{shbrg})), opposed to the time-reversal symmetric $p+ip$ Hamiltonian. 
It follows that it is possible to diagonalise the above schematic Hamiltonian by product 
wave functions of generalized pair creation operators, the so-called Bethe Ansatz states. 
This solution of the $p_x + i p_y$ Hamiltonian was pioneered by Iba\~{n}ez et al. \cite{ibanez:2009}
and further studied by Rombouts et. al.\cite{rombouts:2010} and Dunning et. al.\cite{dunning:2010}. 
The latter serves as a comprehensive article about the $p_x + ip_y$ pairing Hamiltonian and related integrable models.
The free parameters of the ansatz wave functions (the so called RG variables) are determined through the solution
of a system of $N$ RG equations where $N$ is the number of active pairs in the system.
This system of equations is highly non linear and solving it for arbitrary excited states and a realistic number of pairs
and single-particle levels has been a subject of active research \cite{dunning:2010,rombouts:2010,marquette:2012,araby:2012,faribault:2011}. One of the main difficulties of solving the RG equations
is the circumvention of singular points, also called critical points. These singular points arise
when two or more RG variables become equal, and lead to singularities in the RG equations.
\par
The Hamiltonian in Eq.~(\ref{bham}) has an interesting phase diagram: because of the non-zero rotational order the ground state exhibits a quantum phase transition
between qualitatively different superfluid states \cite{rombouts:2010,lerma:2011}. The ground-state energy shows a corresponding nonanalyticity,
as opposed to $s$-wave pairing of which it is well understood that by increasing the interaction strength there is a crossover 
(and not a quantum phase transition), between a weak-coupling Bardeen-Cooper-Schrieffer \cite{bardeen:1957} (BCS) and a quasimolecular Bose-Einstein condensate (BEC) phase \cite{chen:2005}.
In the case of the $p_x + ip_y$ spinless fermion pairing Hamiltonian, this quantum phase transition is only present for 
sufficiently dilute gases $\rho < \frac{1}{2}$ with $\rho$ the fermion density.
The transition itself is continuous, third order and occurs at the so called Read-Green line, 
defined as the points in the phase-diagram where the chemical potential $\mu$ vanishes and BCS mean-field theory predicts a gapless excitation spectrum \cite{ibanez:2009}. The Read-Green line separates the weak pairing regime
from the strong pairing regime. The fingerprint of the Read-Green line is clearly visible in the spectrum of finite systems (see section(\ref{exstates})). 
Another interesting line in the phase diagram at weaker interaction constant is the Moore-Read line where the total energy equals zero,
because all the RG variables collapse to zero, giving rise to a boson-like condensate of equal generalised pairs.
The condensation of all distinct generalised pairs into a power of equal generalised pairs is reminiscent of a bosonic state. 
At stronger interaction constant a second regime occurs, 
the so called `condensate regime';  where a number of RG variables collapse to zero at particular interaction constants.
The Moore-Read line is a special case of this dynamics where \emph{all} the RG variables collapse to zero.
In contrast with the Read-Green line there is no quantum phase transition at the Moore-Read line \cite{rombouts:2010}.
A particular technique that can be used to get more insight into the dynamics of the system and the phase diagram is 'bosonization'.
The process of bosonization maps the hard-core bosons present in the system (RG variables) adiabatically onto real bosons. With this method it is also possible
to investigate the effects of the Pauli principle on the system, because it allows us to gradually reintroduce the Pauli principle.
This technique has already proven its value for the reduced BCS model \cite{debaerdemacker:2012b,sambataro:2007}.
\par 
The goal of this paper is to extend the results of Iba\~nez \cite{ibanez:2009} and Rombouts \cite{rombouts:2010}, employing a new view to 
the phase diagram by linking the eigenstates to associated bosonic states of the Tamm Dancoff Approximation (TDA), by deforming the quasi-spin algebra. This technique,
introduced in a study \cite{debaerdemacker:2012b} of the collectivity of the reduced BCS model, can serve as an RG solver, in addition 
to existing methods \cite{faribault:2011,marquette:2012}. The method is computationally stable and fast. In essence, we avoid the singular
points by linking the solution of the $N$ RG equations to the solution of one non-linear secular TDA equation, which is easily solvable. 
It gives straightforward solutions in the limit of strong and weak interaction constant. 
In the limit of intermediate interaction constant the situation is more complex, but obtaining all solutions in this regime
remains possible, even for large systems.
\par
In the following section~(\ref{shbrg}) we introduce the basic notions and terminology of the hyperbolic RG model. To be
self-contained, we show 
the link with the $p_x + i p_y$ pairing Hamiltonian and derive the nonlinear RG equations.
The concept of the quasi-spin pseudo deformation parameter is introduced in
the proceeding section(\ref{ovlap}), and the connection with collective and bosonic states in the Tamm-Dancoff approximation (TDA) are discussed
In section \ref{sresults} we discuss a number of different regimes of the $p_x + i p_y $ Hamiltonian.
We start with the discussion of the infinite interaction regime for which we derive a symmetry with the finite interaction regime.
We recall and derive some results for the special points of the phase diagram: the Moore-Read and Read-Green line, which define the boundaries of the
`condensation regime'. 
Then we investigate the associated bosonic states of the ground state of a spinless Fermigas with $p_x + i p_y$ pairing interaction living on a two-dimensional disk \cite{rombouts:2010}. 
Some interesting shifts of the associated TDA states occur when the interaction constant is varied in particular when the system crosses the Moore-Read line.
We calculate the overlaps of the RG ground state with a selection of TDA states to improve our understanding of the three different regimes: the weak pairing regime,
the condensation regime, and the strong pairing regime.
To finish this section, we discuss the properties and peculiarities of a positive interaction constant.
At the end, excited states are discussed. We depict from a small system the RG variables of all the fully paired states in the complex plane.
The availability of the entire spectrum of modestly sized systems,
makes it possible to investigate the reminiscence of the Read-Green line for finite-size systems.
A pattern for the TDA state that connects to the first excited state at the Read-Green line is found.
\section{The hyperbolic Richardson-Gaudin model \label{shbrg}}
The families of integrable Richardson-Gaudin models have their roots in a generalised Gaudin algebra \cite{gaudin:1976,ortiz:2005}
which is based on the su(2) algebra of the quasi-spin operators \cite{talmi:1993}.
The generators of su(2) with spin representation $s_j$ such 
that $\langle S_j^2\rangle$ = $s_j(s_j+1)$ are given by:
\begin{eqnarray}
 S_j^0 &= & \frac{1}{2} \biggl( \sum_{m=-j}^{j} \da{c}_{jm} c_{jm} -\frac{\Omega_j}{2} \biggr), \quad S^+_j = \sum_{m > 0}^{j} \da{c}_{jm} \da{c}_{j\bar{m}}, \label{anop} \\
S_j^- & = & \left( S^+_j \right)^{\dagger},
\end{eqnarray}
with $\da{c}_{jm}$ an operator creating a fermion in single-particle state $jm$, with $m$ the projection of the $\Omega_j =2j+1$ degenerate level $j$, and $j\bar{m}$ denotes the time reverse of $jm$.
These operators span the standard su(2) algebra which can be straightforwardly deduced
from the anticommutation relations of the fermion creation and annihilation operators \cite{talmi:1993}.
\begin{equation}
 \left[S_i^0, S_j^{\dagger}\right] = \delta_{ij} S_j^{\dagger}, \quad \left[S_i^{0},S_j\right] = - \delta_{ij} S_j, \quad \left[S_i^{\dagger},S_j\right] = 2 \delta_{ij}S_j^0 \label{su2comr}
\end{equation}
Each su(2) copy is associated with a single-particle level $i$.  The irreducible representations (irreps) are given by
\begin{equation}
 \ket{s_i , \mu_i} = \ket{\tfrac{1}{4} \Omega_i - \tfrac{1}{2} v_i , \tfrac{1}{2} n_i - \tfrac{1}{4} \Omega_i},
\end{equation}
where $v_i$ stands for the seniority (the number of unpaired fermions) of the $i$ th level and $n_i$ is the number of fermions present in the $i$th level.
For doubly degenerate levels ($\Omega = 2$), there are only two distinct irreps: $s_i = 0$ or $s_i = \frac{1}{2}$, corresponding
respectively with seniority $v_i =1$ or $v_i=0$, which are commonly referred to as `blocked' or `unblocked' levels.
An RG integrable model is defined by $L$ Hermitian, number-conserving, and mutually commuting operators with linear and quadratic terms
of $L$ copies of su(2) generators.
\begin{equation}
 R_i = S_i^0 - 2 \gamma \sum_{j \neq i}^L \biggl[ \frac{ X_{ij}}{2} \left(\da{S}_jS_j+S_i\da{S}_j \right)+ Z_{ij} S_i^0S_j^0 \biggr] \label{intmotion}
\end{equation}
The number-conservation symmetry
is very useful because we only need to search in Hilbertspaces with a fixed particle number
to find the eigenstates of the $R_i$ operators, which reduces the complexity of the problem significantly.
Following Gaudin \cite{gaudin:1976} it is now possible to find conditions for the $X$ and $Z$ matrices so all the $R_i$ operators commute mutually. 
There are two families of conditions, the rational and hyperbolic families respectively. The rational model has the conditions,
\begin{equation}
 X_{ij} = Z_{ij} = \frac{1}{D_i^2 - D_j^2},
\end{equation}
whereas the hyperbolic model is represented by
\begin{equation}
 X_{ij} = 2\frac{D_iD_j}{D_i^2-D_j^2} , \quad Z_{ij} = \frac{D_i^2 + D_j^2}{D_i^2-D_j^2}.
\end{equation}
Any linear combination of the $R_i$
operators with the $X$ and $Z$ matrices fulfilling one of the above conditions gives rise to an integrable model. It is possible to construct a schematic $p_x + ip_y$ pairing Hamiltonian out of the above operators with the $X$ and $Z$ matrices fulfilling the hyperbolic conditions, 
\begin{eqnarray}
 \hat{H} &=& \lambda \sum_i D_i^2 R_i \\
\text{with } &\lambda& = \frac{\eta}{ 1 + 2 \gamma \left( 1- N \right) + \gamma \left( L - \sum_i v_i \right)}, \label{lambdapar}
\end{eqnarray}
where $\gamma$ is a parameter proportional to the interaction constant $g = - 2 \lambda \gamma$, and
$N$ the number of pairs.  After some straightforward algebraical 
calculations and subtraction of the diagonal term $g \sum_i \mathbf{S}_i^2 D_i^2$,
the following Hamiltonian appears:
\begin{equation}
\hat{H}_{fac} = \eta \sum_{i=1}^L D_i^2 S^0_i + g \sum_{ij =1}^L D_i D_j^* \da{S}_i S_j. \label{hfac}
\end{equation}
The link with the $p_x + i p_y$ Hamiltonian in eq.(\ref{bham}) is made by redefining $D_i = \frac{k_x - ik_y}{\sqrt{2m}} e^{i\phi}$ and $g = \frac{-G \eta}{2}$. The phase factor $\phi$ is
chosen such that $D_i$ is real and the residual phase factor is absorbed in the corresponding pair creation and annihilation operators (\ref{anop}) without affecting the su(2) quasi-spin algebra.
Since `blocked' levels (seniority $v_i =1$) do not contribute to the pairing interaction, we focus on a full seniority zero space, or 
equivalently, the fully paired space. So the number of active levels $L_c = L - \sum_i v_i$ equals $L$ in our examples.

\begin{figure}[htb!]
\includegraphics[width =8.5cm]{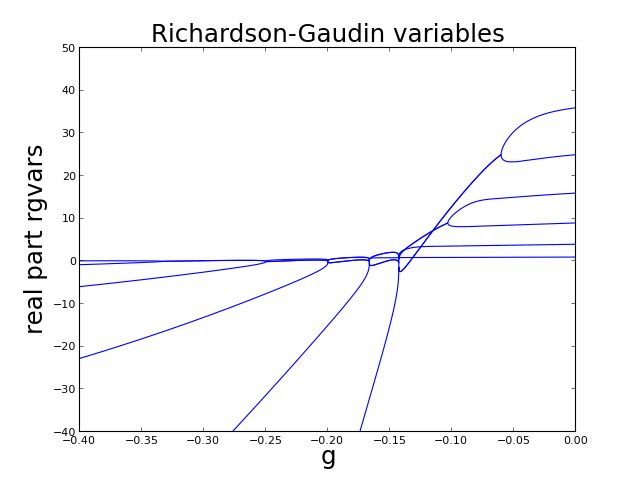}
 \includegraphics[width = 8.5cm]{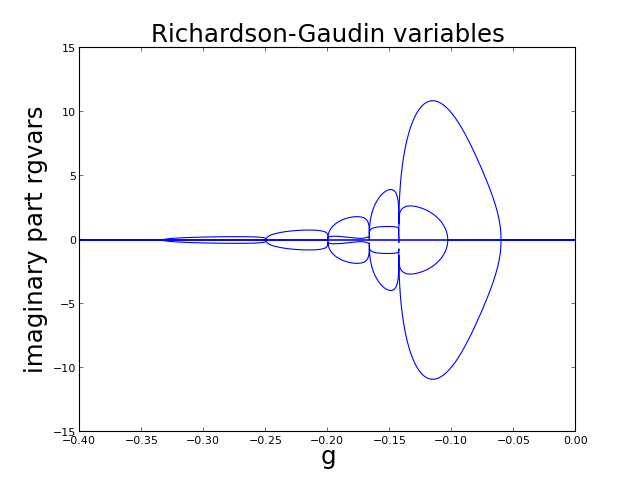}
 \caption{For a system with 12 doubly degenerate single-particle levels
 occupied by 6 pairs, and $D_i = i$, we depict: the real part of the RG variables and the imaginary part of the RG variables in function of the interaction constant $g$.
 Note the qualitative differences between the RG variables of the factorisable Hamiltonian depicted here, and
 those of the rational picket-fence model (cfr. Fig. 1 in Ref [\citenum{debaerdemacker:2012b} ]).
\label{figrgpath}}
 \end{figure}

The Hamiltonian (\ref{hfac}) is built out of $L$ integrals of motion of the hyperbolic RG model.
It follows that the Hamiltonian, the $L$ integrals of motion $R_i$ and the $z$ component of the total quasi-spin, $S_z = \sum_{i =1}^L S_i^z$ have 
a common eigenbasis.
The eigenstates are parametrised by the ansatz \cite{dunning:2010}
\begin{equation}
\ket{\psi} = \prod_{\alpha = 1}^N \da{K}_{\alpha} \ket{\theta} \label{rgstate}
\end{equation}
with $\da{K}_{\alpha}$ a generalised pair creation operator defined as
\begin{equation}
 \da{K}_{\alpha} = \sum_{k=1}^L \frac{D_k \da{S}_k}{\eta D_k^2- E_{\alpha}}. \label{rgvar}
\end{equation}
The state (\ref{rgstate}) is an eigenstate of $\hat{H}_{fac}$ if 
the parameters $E_{\alpha}$ are solutions of the following system of equations \cite{ortiz:2005}
\begin{multline}
 1+2g \sum_{i=1}^L \frac{D_i^2 s_i}{\eta D_i^2 - E_{\alpha} } \\
 - 2 \frac{g}{\eta}\sum_{\beta \neq \alpha}^N \frac{E_{\beta}}{E_{\beta}-E_{\alpha}} = 0, \quad \forall \alpha = 1 \ldots N \label{rgeq}
\end{multline}
The above equations are the RG equations for the $p_x + ip_y$ pairing Hamiltonian.
The total energy of the eigenstate is given by:
\begin{equation}
 E = \sum_{\alpha=1}^N E_{\alpha} - \eta \sum_{k =1}^L D_k^2 s_k \label{totenergy}
\end{equation}
The system of equations as described in eq.(\ref{rgeq}) is equivalent with the RG equations in \cite{rombouts:2010}, with
the definition $D_i = \sqrt{\eta_i}$, $g= -G$, and a rearrangement of $E_{\alpha}$ in the numerator of the 
third term of eq.({\ref{rgeq}}). We have opted for the form in eq.(\ref{rgeq}) for numerical stability, because the constant number 1 in (\ref{rgeq}) acts as a 
reference point for the solver, as opposed to the form  in \cite{rombouts:2010} where the RG variables have an attractor at infinity. 
Remark that the RG equations are ill-defined for $\eta = 0$, however it is possible to make a connection with a $\eta \neq 0$ state see section (\ref{etaosection}).
Furthermore the $\eta = 0$ state is already extensively discussed by \cite{pan:1998,balantekin:2007}. The path of the real and imaginary part of the RG variables
of a toy model, with 12 doubly degenerate levels, and equidistant $D_i = i$, occupied by 6 pairs, in function of the interaction constant is depicted in Fig. \ref{figrgpath}.
\par
An aspect of the RG models not much touched upon is the evolution of the integrals of motion see eq.(\ref{intmotion}).
If the solution of eq.(\ref{rgeq}) is obtained, and the ground state $\ket{\psi} = \prod_{\alpha = 1}^N \da{K}_{\alpha} \ket{\theta}$ is constructed.
It is possible to calculate the integrals of motion corresponding to a particular eigenstate. Acting with $R_i$ on an eigenstate 
of the factorisable interaction Hamiltonian eq.(\ref{hfac}), yields the following eigenvalue:
\begin{equation}
 r_i = s_i \left(-1 - 2 \gamma \sum_{k\neq i}^L Z_{ik} d_k -2 \gamma \sum_{\beta = 1}^N Z_{\beta i} \right)
\end{equation}
with $Z_{\beta i} = \frac{ \frac{E_{\beta}}{\eta} + D_i^2}{\frac{E_{\beta}}{\eta} - D_i^2} $ and
$-2 \gamma = \frac{1}{\frac{\eta}{g} + (1-N) + \frac{L_c}{2}}$. The $E_{\beta}$ are the RG
variables of the eigenstate. 
A remarkable fact is that the integrals of motion associated to particular eigenstates
exhibit singularities at particular $g$.
The Hamiltonian (\ref{hfac}) uniquely defines a set of conserved charges $R_i$ (Eq. (\ref{intmotion})) via the definition of $X_{ik}$, $Z_{ik}$, and the parameters ($\lambda,\gamma$). 
As such, the eigenvalues $r_i$ of $R_i$ exhibit singularities for those values of $g$ where $\gamma$ becomes singular. 
Nevertheless, the eigenvalues of the Hamiltonian contain no traces of these singularities because they cancel exactly by construction via $g=-2\lambda\gamma$.
 

\section{Collective and pseudo deformed states \label{ovlap}}
The eigenstate (\ref{rgstate}) is a product state of generalised pair creation operators $\da{K}_{\alpha}$ (\ref{rgvar}). Opposed
to the constituent fermions, the generalized pair creation generators $\da{K}_{\alpha}$ commute, and are
therefore commonly referred to as `hard-core'  bosonic states. The product wave structure is reminiscent of bosonic 
approximations, such as the Random Phase Approximation (RPA) and $pp$-Tamm Dancoff Approximation (TDA) \cite{rowe:2010,ring:1980}.
Recent investigations on the relation between the $pp$-TDA and the rational RG model \cite{sambataro:2007,debaerdemacker:2012b} have shown
that a one-to-one correspondence is possible between the bosonic-like TDA states and the Bethe ansatz states of the 
rational RG model, either by calculating overlaps \cite{sambataro:2007},
or a pseudo deformation of the algebra \cite{debaerdemacker:2012b}. The ground state of the reduced BCS Hamiltonian in the strong interaction regime has a clear-cut connection
to a condensate of the collective TDA eigenmode, whereas the weak-interaction regime corresponds to a regular filling of the TDA eigenmodes, as dictated by the Pauli principle.
The one-to-one correspondence in the strong interaction limit is particularly remarkable because it is well established that the RG variables in the strong interaction limit are distributed
along an arc in the complex plane, which is not a condensate of equal generalised pairs. In contrast to the rational model, the hyperbolic model supports a fully condensed state
at the Moore-Read line (and fractionally condensed states). Therefore it is of interest whether a similar picture as in the rational model applies for the hyperbolic model.
The basic idea behind the TDA is that it approximates the interacting system as a simple product state of single excitation eigenmodes of the pairing
Hamiltonian (see eq. (\ref{singlestate})).
In the next subsections we elaborate on the method that is used to link those bosonic states with the `hard-core' bosonic states which 
are the $N$ pair eigenstates of the $p_x + ip_y $ pairing Hamiltonian.
This is done by adiabatically increasing the degeneracy of the levels to infinity by means of a deformation parameter in the algebra of the 
factorisable Hamiltonian, linking the collective TDA states adiabatically
with the eigenstates of the $p_x + ip_y$ Hamiltonian.
The method turns out to be a very efficient solver of the highly singular system of eq.(\ref{rgeq}). Even for some hundreds
of pairs and levels this method stays stable. The only drawback is that in the critical regime corresponding
to medium interaction constants the
combination of TDA solutions which will lead to a solution of the Hamiltonian is not known a priori.
\subsection{TDA states}
The elementary eigenmodes of the $pp$-TDA are determined by the 1-pair excitation eigenvalue equation.
\begin{equation}
\hat{H}_{fac} \sum_{i =1}^L Y_i \da{S}_i \ket{\theta} = E \sum_{i =1}^L Y_i \da{S}_i \ket{\theta} \label{singlestate}
 \end{equation}
This equation is exact for the $N=1$ pair system, and therefore has the Bethe Ansatz eigenstate eq.(\ref{rgstate}) with $E_{TDA}$ as the solution of the
RG equation for $N=1$. 
\begin{equation}
 1 + \frac{g}{2} \sum_i \frac{ D_i^2 \Omega_i}{\eta  D_i^2 - E_{TDA}} = 0. \label{TDAeq}
\end{equation}
which  is also commonly referred to as the secular TDA equation. This equation has a geometric interpretation \cite{ring:1980}; there are $L-1$ real solutions
bound between the successive poles $\eta |D_i|^2$ ($i = 1 \ldots L$) and one unbound solution below $\eta |D_1|^2$, also called the `collective' TDA solution.
Each solution defines a TDA eigenmode, so a general TDA state can be built by picking $N$ eigenmodes out of the $L$ elementary (repetition is possible).
A TDA state can be written as:
\begin{equation}
\ket{\psi_{TDA}}=\prod_{i = 1}^N \left(\sum_{k=1}^L \frac{D_k \da{S}_k}{\eta D_k^2- E_{TDA_i}} \right) \ket{\theta}, \label{tdastate}
\end{equation}
which is structurally equivalent to the Bethe Ansatz state (\ref{rgstate}), but instead of using the RG variables as pair energy parameters, the
energy of the TDA eigenmodes are used. \par
The physical interpretation of eq.(\ref{tdastate}) is a state of $N$ 1-pair excitations with 
no correlations between the pairs. If the pair creation and annihilation
operators of eq.(\ref{hfac}) would have bosonic commutation relations, the above state would be an exact eigenstate of eq.(\ref{hfac}).

\subsection{Pseudo-deformation}
The pseudo deformation of the quasi-spin algebra provides a convenient means to adiabatically  connect the exact RG Bethe Ansatz states with the bosonic TDA states. 
The algebra is given by \cite{arrechi:1972,debaerdemacker:2012b}
\begin{eqnarray}
 \com{S_i^0}{\da{S}_j} &= & \delta_{ij} \da{S}_j, \quad \com{S_i^0}{S_j} = -\delta_{ij} S_j, \label{c1}\\
 \com{\da{S}_i}{S_j} &= & \delta_{ij} \left(\xi \hat{n}_j-\frac{1}{2}\Omega_j \right) \nonumber \\ 
 & =& \delta_{ij} \left(\xi 2S^0_i+\left(\xi-1\right)\frac{1}{2}\Omega_j \right), \label{c2}
\end{eqnarray}
where $\xi$ is the pseudo deformation parameter, tuning the Pauli principle
between the full quasi-spin su(2) algebra for $\xi =1$ and a bosonic hw(1) Heisenberg-Weyl ($\xi = 0$) algebra. We employ the term pseudo deformation, because the algebra eq.(\ref{c1},\ref{c2})
is transformable to a genuine su(2) algebra for $\xi \neq 0$, with irreducible representations labelled by
\begin{equation}
\ket{s_i(\xi) , \mu_i(\xi)} = \ket{\tfrac{1}{4\xi} \Omega_i - \tfrac{1}{2} v_i , \tfrac{1}{2} n_i - \tfrac{1}{4\xi} \Omega_i}.
\end{equation}
The physical picture associated with the pseudo deformed irreducible representations is an opening of the sp orbitals by a factor of $\frac{1}{\xi}$, giving rise to an increased
degeneracy of the orbital, with the possibility to accommodate an arbitrary amount of pairs in the $\xi \rightarrow 0$ limit. Because the pseudo deformed algebra
is eventually isomorphic to a genuine su(2) quasi-spin algebra, the Hamiltonian eq.(\ref{hfac}) remains RG integrable with associated pseudo-deformed RG equations:


\begin{multline}
 1+2g \sum_i \frac{D_i^2 \xi s_i\left(\xi\right)}{\eta D_i^2 - E_{\alpha} } 
 \\ - 2\xi \frac{g}{\eta}\sum_{\beta \neq \alpha} \frac{E_{\beta}}{E_{\beta}-E_{\alpha}} = 0. \quad \forall \alpha = 1 \ldots N
 \label{rgeqdef}
\end{multline}
It is easily verified that $\xi =1$ gives rise to the original RG equations (\ref{rgeq}),
whereas the $\xi=0$ limit decouples the RG equations into $N$ independent 1-pair excitation equations (\ref{TDAeq}).
To make the connection from the $\xi = 0$ state to the $\xi = 1$ state in which we are interested it is necessary to have the $\xi \ll 1$ limit under control.
This is because putting more than one pair in the same TDA eigenstate will blow up the third term of eq.(\ref{rgeqdef}) at any $\xi \neq 0$.
Fortunately there exists an approximate solution for very small $\xi$ which depends on the collective solutions by making
use of the Heine-Stieltjes connection \cite{guan:2012,stieltjes:1914}. It resolves the divergences by adding an imaginary part to the collective solutions
associated to sp levels that are occupied by more then one pair (see appendix(\ref{nclim})).
\begin{equation}
  E_{\alpha}^{\nu}\left(\xi\right) \approx E_{\alpha}\left(0\right) - i \sqrt{\frac{2E_{\alpha}\left(0\right)}{\eta a}} z^{\nu}  \quad \xi \ll 1 \quad \forall \nu = 1\ldots n
  \label{esxi}
\end{equation}
With $z_{\alpha}^{\nu}$ the $\nu$-th root of the `physicists'  Hermite polynomials $H_{n}\left(z\right)$, $a_{\alpha}$ given
by: $a_{\alpha}= \frac{1}{2} \sum_i \frac{D_i^2 \Omega_i}{\left(\eta D_i^2 - E_{\alpha}\left(0\right)\right)^2}$ and $\nu \in [1,\ldots,n]$ where $n$ is the number
of pairs associated with a collective solution $E_{\alpha}\left(0\right)$.
Eq.(\ref{esxi}) contains a lot of information about the underlying structure of the RG variables. By choosing a TDA distribution 
corresponding to an eigenstate of eq.(\ref{hfac}), eq.(\ref{esxi}) answers immediately the question if a RG variable will be complex or real when
the system is not in the 'condensate' regime.
So the imaginary 
character of a RG variable depends on the roots of the Hermite polynomials, the number of pairs associated to a TDA level $n$ and the sign of the corresponding
TDA solution $E_{\alpha}(0)$. 
As an example, in the weak interaction limit, where the structure of 
the system can be regarded as a simple filling of the fermi sea with hard-core bosons, with only
doubly degenerate levels, all the solutions are real because $n = 1 \; \forall \alpha$, and the 
roots of the Hermite polynomial of first order are zero. In the strong interaction
limit all pairs are associated with the lowest TDA eigenmode which is negative for $g\rightarrow -\infty$.
Therefore we see from eq.(\ref{esxi}) that 
all the RG variables are real. So for a set of doubly degenerate levels we can
only have complex RG variables at intermediate interaction constant. 

\subsection{RG solver}
The solution method described above for eq.(\ref{rgeqdef}) can be used as an efficient solver for the hyperbolic RG equations.
The absence of correlations in the TDA states reduces the computational complexity of the problem significantly, because only one equation (\ref{TDAeq}) needs to be solved as opposed to $N$ coupled equations(\ref{rgeq}). 
This is the key idea behind the RG solver. The uncorrelated system is solved and then the full pairing problem is retained by adiabatically reintroducing the Pauli principle. 
We label the TDA eigenstates with a partitioning of $N$ out $L$ integers. This means that the state is labelled by vectors of integers $(\nu_1, \nu_2, \ldots, \nu_l)$ with length $L$ and $\nu_i = 0,\ldots, N$, with 
the additional constraint that $\sum_{i=1}^L \nu_i = N$. 
Two interesting cases are the fully collective case $(N , 0, \ldots ,0)$ corresponding with the ground-state in the strong-interaction regime, and $\left(\underbrace{\frac{\Omega_1}{2}, \ldots, \frac{\Omega_{n-1}}{2}, \nu_n}_{n \left(\nu_n < \frac{\Omega_n}{2}\right)},\underbrace{0 , \ldots , 0}_{L-n} \right)$
corresponding with the ground-state in the weak interaction regime, which have proven to play a pivotal role in the rational case \cite{debaerdemacker:2012b}.
When the interaction constant approaches zero, the TDA collective states and the actual physical eigenstates become equal to a filling of pairs of the lowest sp levels
up to the Fermi surface.
This is because the pairing interaction behaves as a very small perturbation on the sp levels in that case.
This makes it possible to label a RG eigenstate with the TDA distribution of pairs that connects to that RG state in the weak interacting limit.
The maximum number of pairs that can be associated
to a TDA-solution in the weak interaction regime, is never more then the total pair degeneracy of the corresponding sp level.
In the intermediate interaction regime the RG states connect to TDA states with some eigenmode multiplicities larger than the degeneracy of the corresponding levels, but lower or equal
than the total number of (collective) pairs. 
When the interaction constant becomes stronger the collectivity of the TDA state associated to the RG groundstate increases gradually.
Until the most collective TDA state connects to the RG ground state, in this TDA state all pairs occupy the collective TDA eigenmode.
In the very weak and strong pairing regime it is clear which state connects to the RG ground state. For the intermediate regime this is not the case, and an educated guess
for the TDA start distribution has to be made. An alternative solution method is to obtain a solution in the very weak or strong interaction limit and then changing $g$ with
small steps until the desired interaction constant is reached. Singular points can be circumvented by a continuation of $g$ in the complex plane \cite{dussel:2007}, or reducing the $\xi$ value which
enhances the effective degeneracy of the single-particle levels ($s_i\left(\xi\right) = \frac{1}{4\xi} \Omega_i - \frac{1}{2} v_i $), and therefore it has a softening effect on the singular points. 
Using this approach it is possible to solve systems of hundreds of levels occupied by hundreds of pairs \cite{vanraemdonck:2013}.
In practice we use our method to obtain a solution in a limit where the
TDA distribution for the state of interest is known and then gradually change the interaction constant to the interaction constant of interest. Critical points are circumvented by giving the interaction
constant a small complex phase or deforming the pairing algebra. All the calculations presented in this paper were performed on a standard desktop computer. Results for a system with 256 levels and 128 pairs
were obtained for a full range of the interaction constant in a few hours. If solutions for a full range of the interaction constant
need to be calculated then most of the calculation time is spent in the circumvention of critical points. When critical points are circumvented it is also necessary to check the continuity
of the energy regularly, because the possibility exists that the RG variables jump to a different state. The newton-raphson method was used to solve the pseudo-deformed and normal RG equations.
The proposed method is very fast and stable if the associated TDA distribution is known a priori for the state of interest, for example for the first excited and ground state at the Read-Green point (Section \ref{exstates}). 


\section{Different regimes \label{sresults}}
In this section we investigate first the connection between the $\eta = 0 $ and $\eta \neq 0$ systems. Next we use the tools developed
in the previous section to learn more about the Moore-Read line and the two regimes of which the Moore-Read line is the line of demarcation. 
\subsection{The $\eta = 0$ Hamiltonian \label{etaosection} }
A connection is made between the $\eta = 0$ state and the state with $\eta = -2g$. This is relevant because eq.(\ref{rgeq}) diverges when $\eta \rightarrow 0$.
So by having a method to solve the $\eta \neq 0$ case we are able to generate the solutions of the $\eta = 0$ case.
The Bethe ansatz solution of the $\eta = 0$ state was first explored by Pan et. al. \cite{pan:1998} and later by 
Balantekin et. al. \cite{balantekin:2007} who explored some symmetry
properties of the Bethe-ansatz equations. Two separate sets of Bethe-ansatz equations were found, solutions of the first 
set were zero and the solutions of the other set were not constricted to zero. \par
\begin{figure}[htb!]
 \includegraphics[width = 8.5cm]{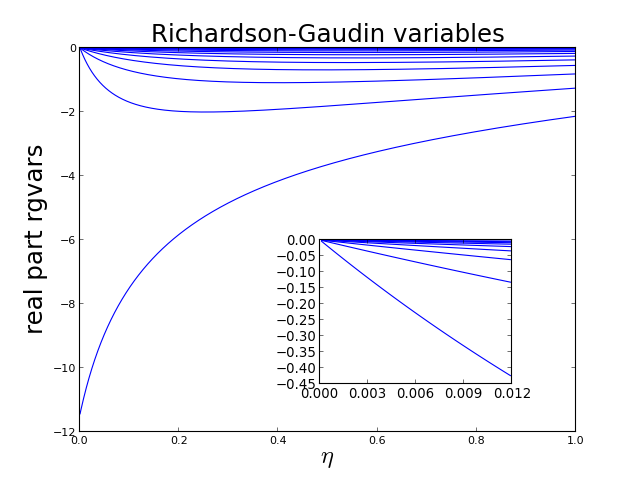}
\caption{\label{eta=0} Depicted is the evolution of the real part of the RG variables of the ground state when
$\eta$ evolves from zero to one for a system with level parameters as described in Table \ref{tabsplev} occupied by 10 pairs and $g = -0.075$.
Note that the RG variables remain real during the entire trajectory of $\eta$ because the system remains in the strong pairing regime.}
\end{figure}

Suppose that we have found the eigenstate of the factorisable Hamiltonian $\ket{\psi} = \prod_{\alpha} \da{K}_{\alpha} \ket{\theta}$ with $\eta = -2 g$ for $N$ pairs. 
Then we can write the Hamiltonian as $\hat{H} = -2 g K^0_D + g \da{K}_D K_D \equiv g K_D \da{K}_D$, with 
\begin{equation}
 \da{K}_{D} = \sum_k D_k \da{S}_k \quad K_{D} = \sum_k D_k^* S_k \quad K_{D}^0 = \sum_k D_k^2 S_k^0.
\end{equation}
Note that $\da{K}_D$ can not be written in the conventional $\da{K}_{\alpha}$ form (\ref{rgvar}).
By multiplying the eigenvalue equation $\hat{H} \ket{\psi} = E \ket{\psi}$ with $\da{K}_D$, we obtain:
\begin{equation}
 \da{K}_D g K_D \da{K}_D \prod_{\alpha} \da{K}_{\alpha} \ket{\theta} = E \da{K}_D \prod_{\alpha} \da{K}_{\alpha} \ket{\theta}
\end{equation}
So it is clear that the state $\da{K}_D \prod_{\alpha} \da{K}_{\alpha} \ket{\theta}$ is an eigenstate of the Hamiltonian (\ref{hfac}) with $\eta = 0$ and
$N+1$ pairs. At this point the only question that remains to be solved is: \textquotedblleft What accounts for the mismatch in Hilbert space dimensions?\textquotedblright. If 
$L$ denotes the number of levels then the Hamiltonian with $\eta = -2g$ has $\binom{L}{N}$ states in the fully paired space and the Hamiltonian
with $\eta = 0$ has $\binom{L}{N+1}$ eigenstates. The resolution of this seemingly paradox resides in the fact that with $\rho$ lower than half-filling
the extra eigenstates of the system with $\eta = 0$ have zero eigenvalue \cite{pan:1998,balantekin:2007} and these 
extra eigenstates match exactly the number of missing eigenstates in the $\eta = -2g $ case, above half-filling 
the opposite situation occurs, which indicates a symmetry between those states. Another interesting feature is that the RG variables of a particular state with the same energy in both systems
are not equal but add up to the same energy eq.(\ref{totenergy}). See Fig. \ref{eta=0} for a picture that shows the behaviour of the
RG variables as $\eta$ approaches zero for a system with parameters given in Table \ref{tabsplev} at quarter filling.
\subsection{Three regimes at attractive interaction constant}
The RG equations become singular when two or more RG variables are equal as can be seen in eq.(\ref{rgeq}).
More in particular, at the singular points $2s_i +1$ RG variables occupy only one single-particle level $i$ and are therefore equal \cite{dominguez:2006}.
Those singular points correspond to a reordering of the 
corresponding bosonic states in the case of the rational RG model\cite{debaerdemacker:2012b}. 
This is in contrast with the factorisable interaction model, where this is only the case for interaction constants weaker then the Moore-Read point,
as we will show in the next subsection. Another difference with the rational RG model is the occurrence of
the so called `condensate regime' where a number of RG variables
collapse to zero at particular interaction constants: 
\begin{figure}[htb!]
\includegraphics[width =8.5cm]{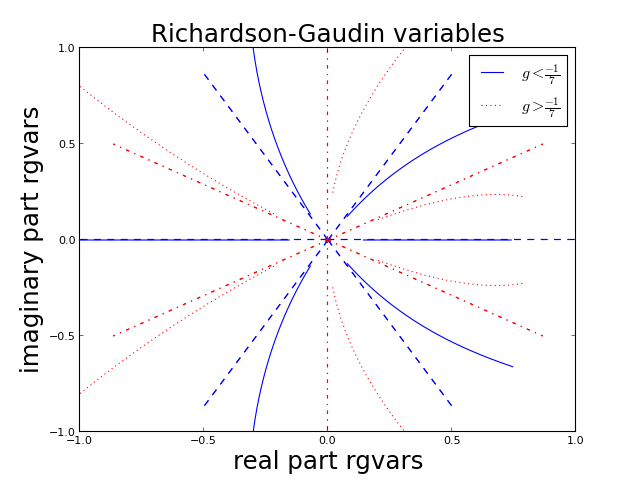}
 \caption{The behaviour of the RG variables in the neighbourhood of the Moore-Read point is depicted for a system with 6 pairs in 12 two-fold degenerate levels, and $\eta = 1$
 (see Fig. \ref{figrgpath}). The evolution of the corners of the two regular hexagons are 
 depicted respectively by a dashed, and a dot-dashed line. The Moore-Read point occurs at $\frac{\eta}{g} = -7$.
\label{mrclose}}
 \end{figure}
\begin{equation}
 \frac{\eta}{g} = 2 q + p -1 - 2 \sum_{k}s_k \label{conpoint}
\end{equation}
with $p$ the number of RG variables which have condensed to zero and $q$ the number of generic non-zero RG variables\cite{rombouts:2010}
(see appendix \ref{conap} for an alternate derivation of this formula).
In the continuum limit, the above formula becomes: $\frac{\eta}{gL} = \rho - 1$ and $\frac{\eta}{gL} = 2\rho - 1$
respectively for $N$ and $1$ condensed pairs with $\rho = \frac{N}{L}$ and $gL$ kept constant when $L,N \rightarrow \infty$. 
It follows that the points in phase space with $N$ and $1$ condensed pairs correspond to the Moore-Read and Read-Green line for finite systems.
The Read-Green and Moore-Read points form the boundaries of the condensation regime. The Read-Green line separates the strong pairing regime and
the condensation regime, the Moore-Read line separates the weak pairing regime and the condensation regime.
The strong pairing regime only exists below or at half-filling, above half-filling the system never exits the condensate regime. 
Around those condensation points it is possible to split up the RG equations in two separate sets in lowest order perturbation theory, a set for the condensed RG
variables and one for the non-condensed RG variables. The dynamics of the condensed
RG variables in the neighbourhood of their condensation points is described by regular polygons and the requirement that the RG variables
need to obey a mirror symmetry with respect to the real axis. To fix ideas, if there are 6 pairs which condense to zero then they approach a condensation point on the corners of a regular hexagon, with all corners in the complex plane.
After the condensation point, an extra RG pair stays real, and only an even number of pairs can become complex, so the RG variables leave the condensation point on a regular hexagon
with two corners on the real axis (see Fig. \ref{mrclose}).
The system that describes the non-zero RG variables in the neighbourhood of a condensation point is given by:
\begin{equation}
\begin{split}
\frac{p+1}{2 E_{\alpha}}& + \sum_i \frac{s_i}{\eta |D_i |^2 - E_{\alpha}} \\& - \sum_{\beta \neq \alpha, \beta = p+1}^N \frac{1}{E_{\beta}  - E_{\alpha}} = 0. \quad \forall \alpha =  p+1, \ldots, N \label{con2}
\end{split}
\end{equation}
Remark that the labelling of the RG variables is arranged so the first $p$ RG variables correspond to
the condensed RG variables and the last $N-p$ RG variables are non-condensed.
The position of the collapsed RG variables in the neighbourhood
of their condensation point is determined by
\begin{equation}
E_{\alpha} = z_0 e^{\frac{2\pi i\alpha}{p}}, \quad \alpha = 1 \ldots p. \label{conpoints}
\end{equation}
$z_0 = |z_0| e^{i \phi}$ has a phase that forces mirror symmetry around the x-axis, e.g. for 6 condensed pairs $\phi = 0$ for $g < g_{con}$ and $\phi = - \frac{\pi}{6}$ for $g > g_{con}$, and $|z_0|$ approaches zero.
The behaviour of the condensed RG variables around their condensation points is only influenced by the other pairs through their number, and the number of pairs which are real.
At the Moore-Read line there are only condensed pairs, and the position of all pairs is determined by eq.(\ref{conpoints}).
(For a derivation see appendix \ref{rgcon}.) In the next two subsections the goal is to gain a better understanding
of the three regimes (weak, strong pairing and condensate regime), by investigating, the RG variables and their associated TDA states. 

\subsection{Connecting the TDA state with the RG ground state}
We apply the machinery developed above on a spinless Fermigas with $p_x + ip_y$ pairing interaction symmetry on 
a disk with a radius of five unit cells in a two-dimensional square lattice of which we found the sp characteristics in \cite{rombouts:2010}, see table(\ref{tabsplev}) for
the sp characteristics.
\begin{table*}[htb!]
\begin{ruledtabular}
\begin{tabular}[c]{l*{13}{c}}
 $|D_i|^2$ &0.04& 0.08 & 0.16 & 0.20 & 0.32& 0.36& 0.40 & 0. 52 & 0.64 & 0.68 & 0.72& 0.80 &1.00 \\
$\Omega_k$ & 4 & 4 & 4 & 8 & 4 & 4 & 8 & 8 & 4 &8 & 4 & 8 & 12\\
$s_k$ & 1 & 1 & 1 & 2 & 1 & 1 & 2 & 2 & 1 & 2 & 1 & 2 & 3 \\
\end{tabular}
\end{ruledtabular}
\caption{\label{tabsplev} Level parameters $\eta_k$ and $\Omega_k$ for a disk with a radius of five unit cells in a two-dimensional square lattice. As can
be found in \cite{rombouts:2010}}
\end{table*}

\begin{table}[htb!]
 \begin{tabular}[c]{c|*{6}{l}}
  \hline $ g $ & $\nu_1$ & $\nu_2$ & $\nu_3$ & $\nu_4$ & \ldots & $\nu_{12}$ \\
  \hline \hline 0.00000 & 2 &2 &2 & 4&\ldots & 0 \\
  -0.01518 & 2 & 2& 6& 0& \ldots& 0 \\
   -0.02329 & 2 & 8& 0& 0& \ldots& 0 \\
    -0.02525 & 6 & 4& 0& 0& \ldots& 0 \\ 
     -0.02550 & 7 & 3& 0& 0& \ldots& 0  \\
      -0.02564 & 8 & 2& 0& 0& \ldots& 0 \\
      -0.02690 & 9 & 1& 0& 0& \ldots& 0 \\
       -0.02750 & 10 & 0& 0& 0& \ldots& 0 \\
 \end{tabular}
 \caption{\label{tabcolr}
 The associated collective states of the ground state of a spinless Fermi gas with $p_x+ i p_y$ pairing
 interaction symmetry in function of the interaction constant $g$. With single-particle
 levels given by table(\ref{tabsplev}). $\nu_n$ corresponds to the 
 occupation of the $n^{th}$ TDA solution. The Moore-Read point is located at $g = -0.03225$.}
\end{table}

\begin{figure*}[htb!]
 \includegraphics[width = 5cm]{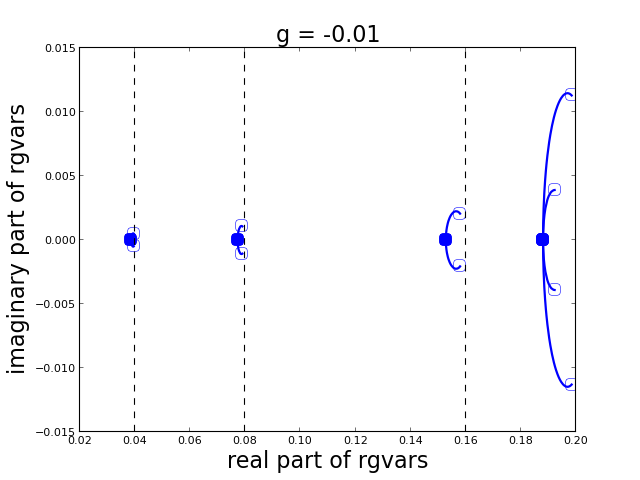}
  \includegraphics[width = 5cm]{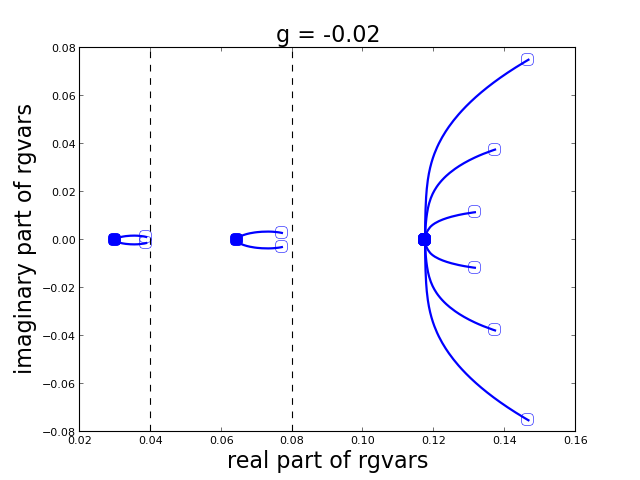}
   \includegraphics[width = 5cm]{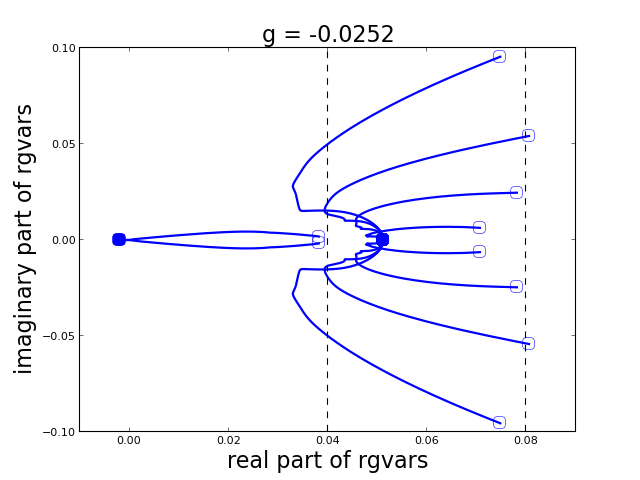}
    \includegraphics[width =5cm]{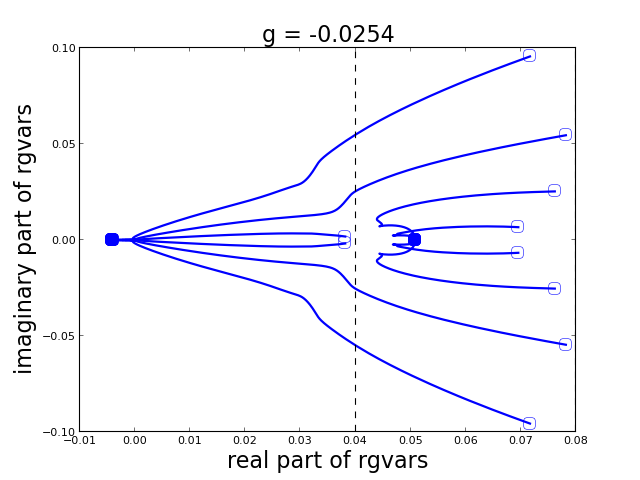}
    \includegraphics[width = 5cm]{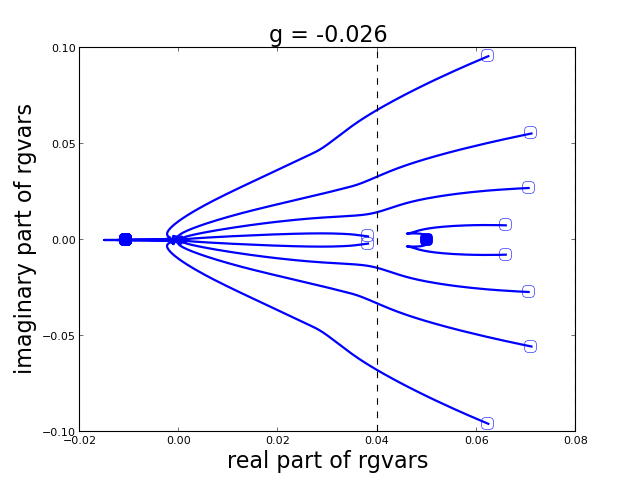}
    \includegraphics[width = 5cm]{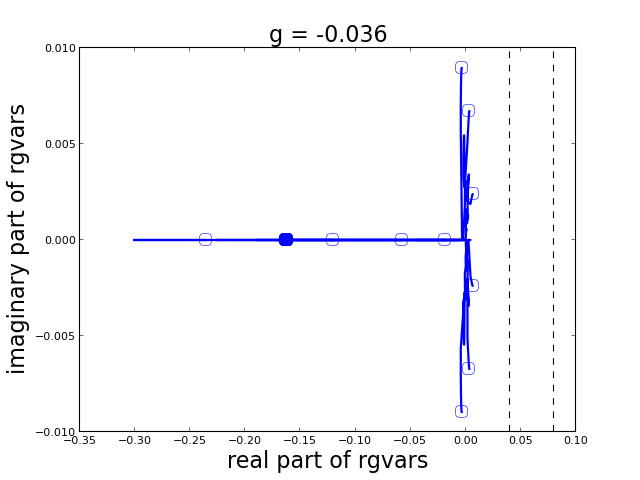}
    \includegraphics[width = 5cm]{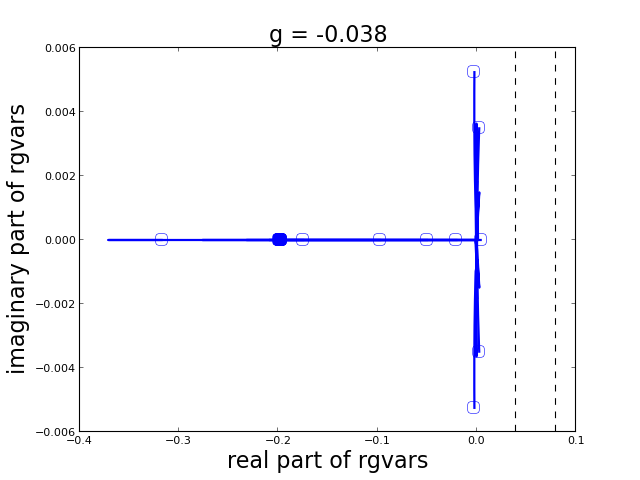}
  \includegraphics[width = 5cm]{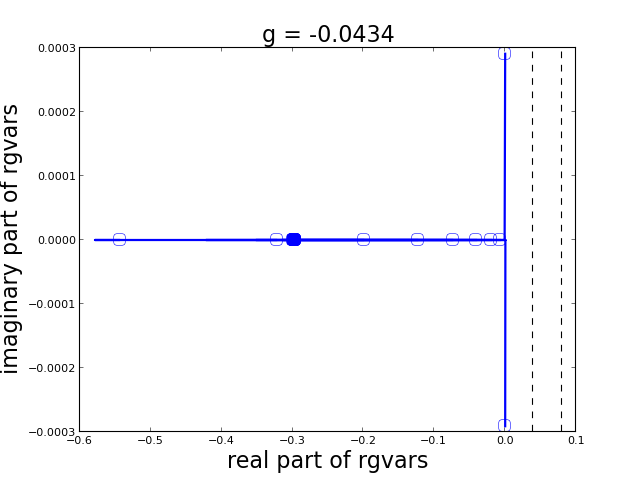}
    \includegraphics[width = 5cm]{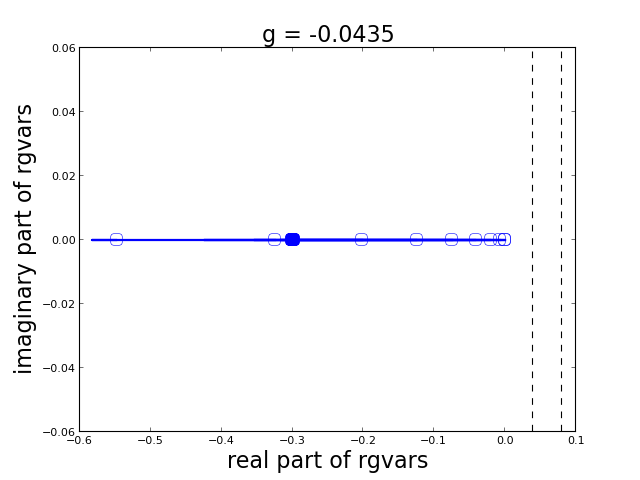}
    \caption{\label{figxipath} The path of the deformed RG variables $E_{\alpha}\left(\xi \right)$ in the complex plane
    for the two-dimensional Fermigas of which the levels are depicted in Table \ref{tabsplev}, for some 
    well chosen values of $g$: $g$ = -0.01, $g$ = -0.02, $g$ = -0.0252, $g$= -0.0254, $g$ = -0.026, 
     $g$= -0.03600, $g$=-0.038, $g$ = -0.0434 , $g$=-0.0435. The path starts from
    the bosonic eigenmodes $\left( E_{\alpha}\left(0\right) \right) = \left(\hbar \omega \right)$ depicted with thick dots and ends at the exact
    RG variables depicted with open dots. The vertical dashed lines indicate the singularities in eq.(\ref{TDAeq}).}
\end{figure*}
If we look at table(\ref{tabcolr}), and Fig. \ref{figxipath}, we notice that the amount of collectivity, as measured by the occupation
of the lowest TDA solutions, gradually increases with stronger interaction constant.
A particularly interesting result is the fact that the most collective TDA state connects to the RG ground state
just before the Moore-Read line where all pairs collapse to zero. However the connection there is not very stable, and this remains during the entire 'condensation' regime.
We have to resort to an imaginary deformation parameter $\xi$ at particular points to make the connection. Outside the 'condensation' regime the connection is stable, and imaginary
deformation parameters are not necessary.
It is also clear that because
of the degeneracy of the sp levels the ground state at low interaction constant
corresponds to a TDA distribution of (22240...0) for 10 pairs. Every single-particle level is able to contain an even number of pairs, so if we turn
the Pauli principle on by increasing $\xi$, the RG variables combine into complex conjugate pairs even at very weak interaction constant, 
as opposed to systems with only two-fold degenerate sp levels where the RG variables are real for small interaction
constants. In that case, only one pair is associated to each TDA eigenmode, and the RG variables can only recombine into complex conjugate pairs if
two neighbouring pseudo-deformed RG variables approach a singularity, and recombine in a complex conjugate pair.
The connection with the (730...) and (910...) state is only present for a very small interval of the interaction constant, and should
be seen as a boundary for a transition of the system of one even state to another.
Recapitulating the findings of this section, we find at small interaction constant a regime for which the ground state gradually connects
to more collective TDA states with increasing interaction constant. The reordering of pairs of the associated bosonic state occur at singular points or in between singular points.
Whenever a singular point occurs there always is a reordering of the associated TDA state. This happens until the Moore-Read line where the TDA-state is in
the most collective form. In the weak pairing regime the connection and the associated TDA states have strong similarities with the reduced BCS Hamiltonian \cite{debaerdemacker:2012b}.
During the 'condensate' regime when the interaction constant fulfils $2N -2 - L_c \geq \frac{\eta}{g} \geq  N-1 - L_c$,
the connection with the most collective TDA state remains but we have to resort to a complex deformation parameter,
until the last condensation point is passed. In the strong pairing regime,
the connection with the most collective TDA state is firmly established. 

\subsection{Overlaps with the collective states}
\begin{figure}[htb!]
 \includegraphics[width = 8.5cm]{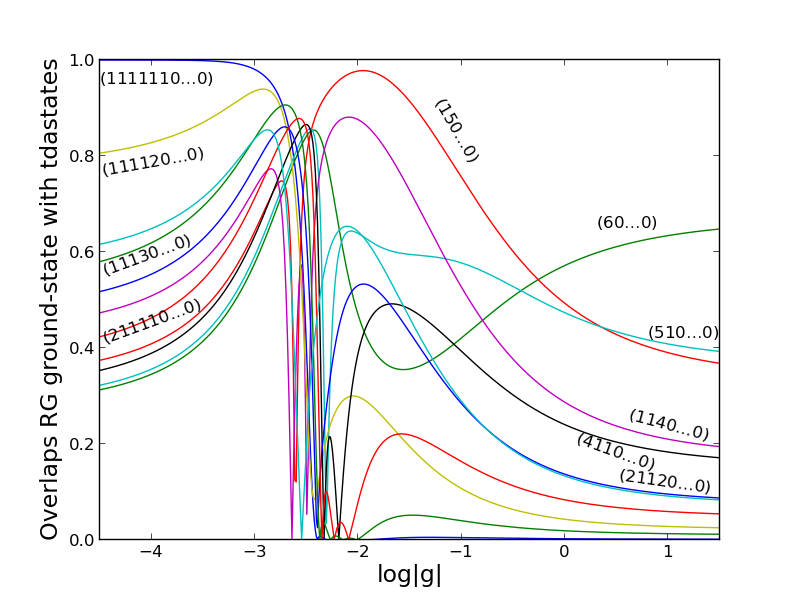}
 \caption{\label{overlap} Depicted are the overlaps of a selected set of bosonic states with the ground state of a system
 with 12 doubly degenerate levels, $\eta =1$, and $|D_i | = 1$ in function of the interaction constant. 
 The bosonic states are labelled according to their TDA eigenmode occupation. The notation is as follows $(\nu_1\nu_20\ldots0)$
 means that the bosonic state is constituted of $\nu_1$ bosons in the TDA state with the lowest $E_{TDA}$ and $\nu_2                                     $ bosons in the first excited TDA state.}
\end{figure}

In this subsection we investigate the overlaps of the ground state of a factorisable interaction Hamiltonian with some selected TDA states
over an entire range of the interaction constant. These overlaps have proven to provide valuable information about the 
RG states and their collective character \cite{sambataro:2007,debaerdemacker:2012b}. Investigations of the overlaps shows that at weak interaction constant the
behaviour of the RG variables resembles that of the reduced BCS Hamiltonian\cite{debaerdemacker:2012b}. After the Moore-Read line
this is not the case anymore.
Fig. \ref{overlap} depicts the overlaps of some well chosen TDA states with the ground state of the system consisting of 6 pairs in 12 doubly degenerate
sp levels (cfr. Fig. \ref{figrgpath}).
We see that for very small interaction constant the overlap of the
RG grounstate with the TDA ground state $\left(1111110\ldots0\right)$ is almost equal to one, as expected. Then there is an intermediate regime 
where some other TDA states with increasing collectivity have the highest overlap with the RG ground state. The interaction
constants where this occurs are the same as the interaction constants where the TDA state that connects to the ground state
changes. Until this point a similar behaviour as in the reduced BCS case is observed. However, the situation alters as the condensation
regime is approached. Here the most collective TDA state $\left(60\ldots0\right)$ goes to a local minimum, while most other states exhibit a maximum in that region.
The TDA state with 1 pair in the lowest TDA solution and 5 pairs in the first excited TDA state has 
the largest overlap, although the most collective TDA state connects to the RG ground state. This
peculiar behaviour starts around the Moore-Read point, so in 
the 'condensate regime' it is no longer true that the TDA state with the highest overlap with the RG ground state connect to the RG ground state according to our scheme. 
The reason for this is that after the Moore-Read point some RG variables that are still complex have 
very small negative real part. The overlap with the $(150\ldots0)$ state is largest here because 5 RG variables
are very close to the 1st excited TDA state and 1 is strongly negative close to the lowest TDA level. The reason why that TDA state does not 
connect to the RG ground state is probably caused by the singularity in eq.(\ref{rgeqdef}) when some RG variables approach zero. Therefore,
all the deformed RG variables have to depart from the lowest TDA solution to connect with the RG ground state of eq.(\ref{hfac}). With increasing interaction constant,
the most collective TDA-state gradually becomes the TDA state with the largest overlap with the ground state of the $p_x+ip_y$ pairing Hamiltonian.
This happens after the condensate regime when all the RG pairs become real.
From then on, the TDA state with the highest overlap with the RG ground state is again the state which connects to the RG ground state, by the pseudo deformation.
However the overlap of the most collective TDA state in the strong interaction regime with the RG ground state is not as prominent as in the reduced BCS case \cite{debaerdemacker:2012b}.
The natural question that occurs is: \textquotedblleft Will the overlap of the most collective TDA state with the RG ground state approach one 
in the limit of very strong interaction constant?\textquotedblright. If we calculate the overlap of the system depicted in Fig. \ref{overlap}
but with $\eta \ll 1$ (which corresponds to the limit of large interaction constant), then we see that in this limit all the overlaps of
the TDA states with the RG ground state have a value around 0.660 and the $(60\ldots0)$ state has the largest overlap with a value of 0.668.
This gives an indication that even at very big interaction constant the overlap of the most collective TDA state with the RG ground state
will never approach one.
This plateau appears to be density dependent, increasing with decreasing density. We conclude that according to the overlaps there are three different regimes: at low
interaction constant a regime that shows similarities with the reduced BCS Hamiltonian and after the Moore-Read point a regime that is significantly different
with a minimum of the overlap of the TDA state which connects to the RG ground state. After the Read-Green line this is restored and the TDA state that 
connects to the RG ground state has the largest overlap again. Opposed to the rational case, there is no consistent isomorphism between the TDA states connecting to the RG ground state
via the pseudo deformation, and the TDA state with a maximal overlap.

\begin{figure*}[htb!]
\subfigure[ The ground state]{
 \includegraphics[width = 4.2cm]{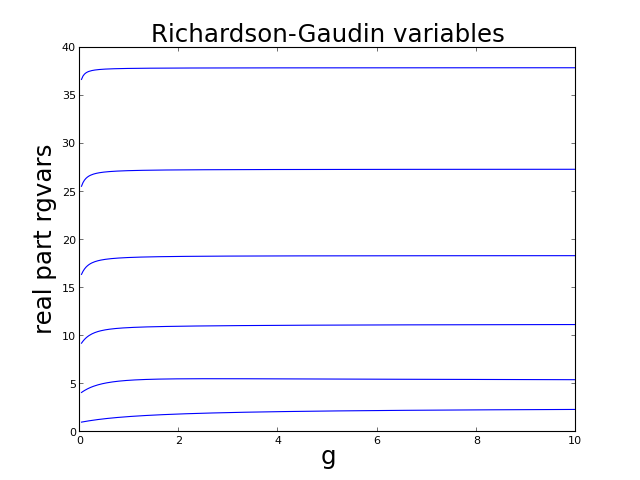} 
  \includegraphics[width = 4.2cm]{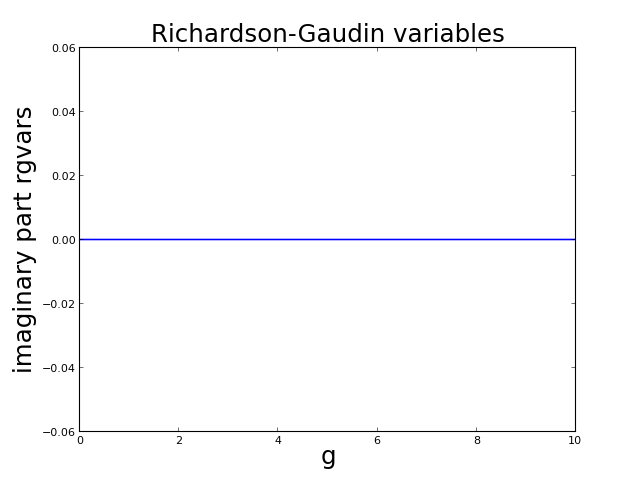}
  \label{6a}}
\subfigure[ The (0 1 1 1 1 1 1 0 0 0 0 0) state]{
 \includegraphics[width = 4.2cm]{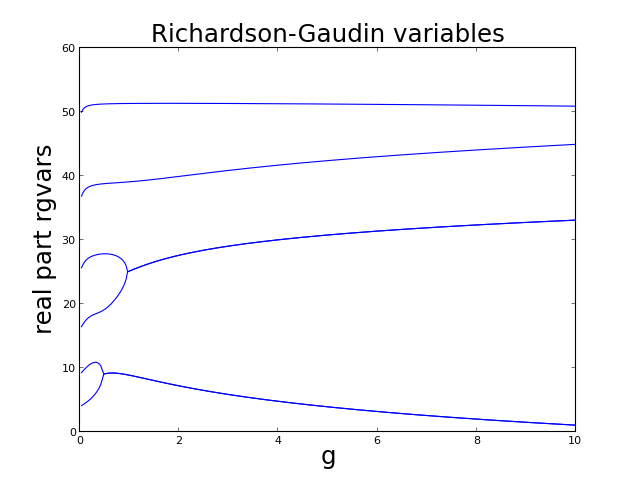} 
 \includegraphics[width = 4.2cm]{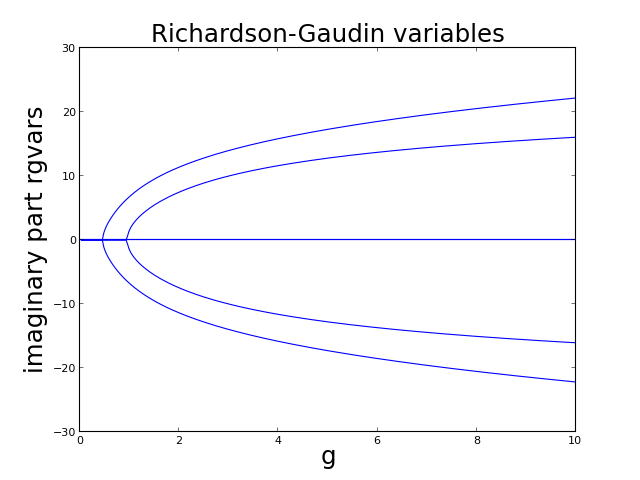} \label{6b}}
\subfigure[ The (0 1 0 1 1 0 1 0 1 0 1 0) state]{
 \includegraphics[width = 4.2cm]{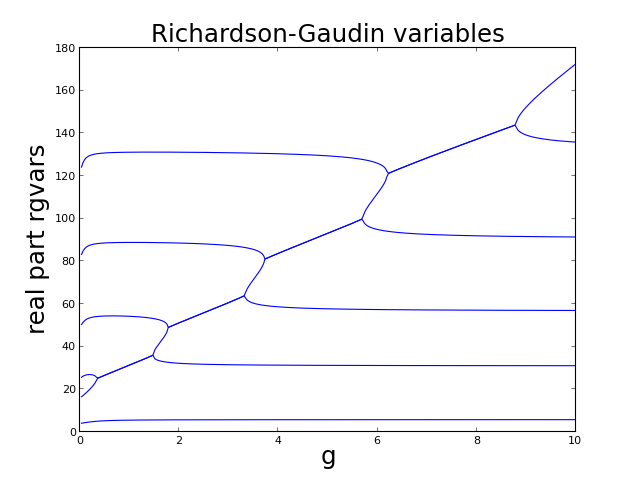} 
 \includegraphics[width = 4.2cm]{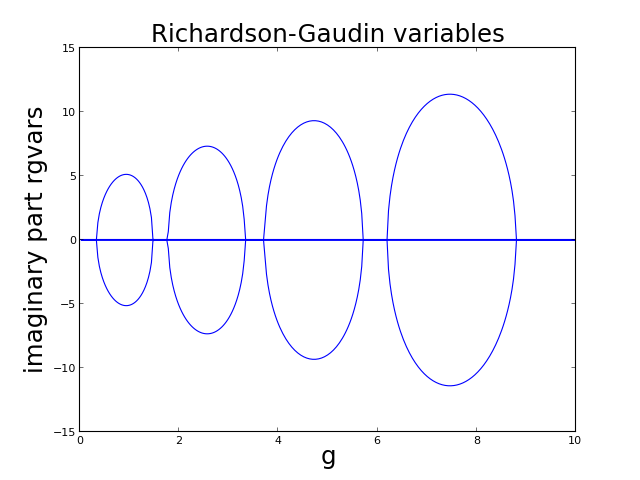}}
\subfigure[ The (1 1 1 0 0 0 0 1 0 0 1 1) state]{
 \includegraphics[width = 4.2cm]{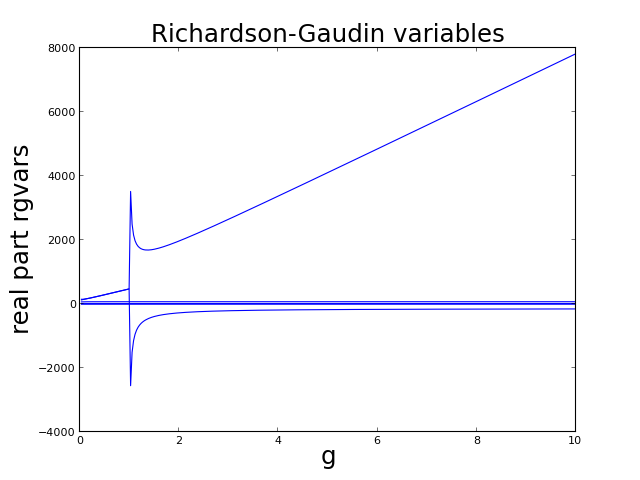}
 \includegraphics[width = 4.2cm]{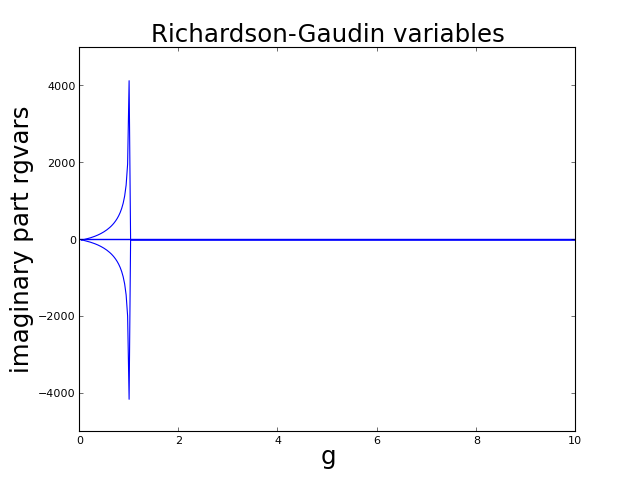}}
\subfigure[ The (0 0 1 1 1 1 0 0 1 0 1 0) state]{
 \includegraphics[width = 4.2cm]{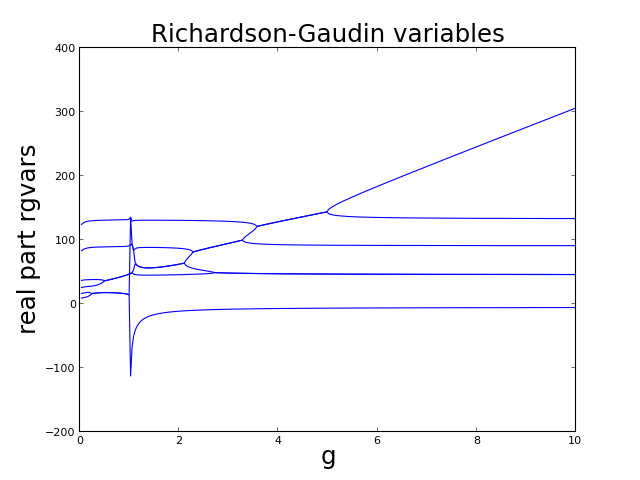} 
 \includegraphics[width = 4.2cm]{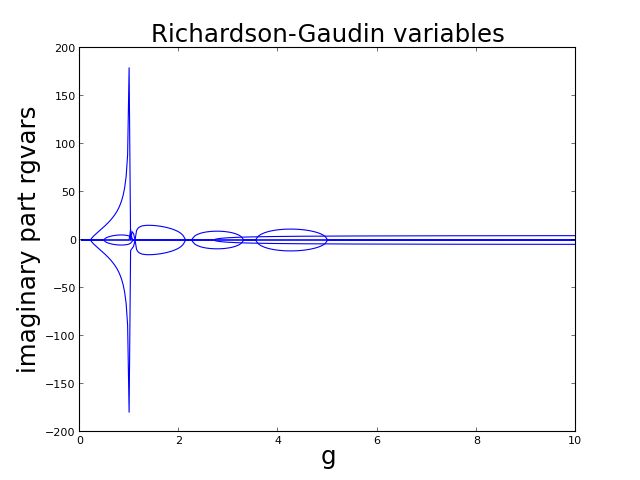}}
 \caption{ \label{poskop724} The real and imaginary part of the RG variables 
 of a system with 12 doubly degenerate sp levels and 6 pairs in function of a positive interaction constant $g$ of some well chosen eigenstates.}
\end{figure*}

\subsection{Repulsive $p$-wave interactions}
There is no mean-field solution in the yet unexplored repulsive case available. So the exact solution method 
presented above offers a unique tool to investigate repulsive $p$-wave interactions. A difference
compared to attractive pairing interaction is that the RG variables now recombine to higher TDA states instead
of lower, as the interaction strength is increased. 
Pairs in isolated sp levels can't recombine,
so the RG variables corresponding to those sp levels that remain real and close to the TDA solution for the entire range of the interaction strength.
There are even start TDA states with neighbouring occupied sp levels
remain real during the entire trajectory, and do not couple to complex conjugate pairs, as is visible in 
the trajectories of the RG variables of the ground-state energy in Fig. \ref{poskop724}a. The RG variables of some excited states 
follow similar trajectories. In general the trajectories of the RG variables for the hyperbolic
RG Hamiltonian with repulsive interaction constant exhibit
three different features. 
\begin{itemize}
 \item A RG variable can remain real during the entire trajectory of the interaction constant, see Fig. \ref{6a}.
 \item Two real RG variables can recombine into a pair of complex conjugate variables by creating a singular point in the trajectory space, after which the complex part gradually increases, see Fig. \ref{6b}. 
 \item Two complex conjugate RG variables can become real again through a sudden jump in complex space and a similar jump in real space. Remark that the jump of the real parts of the RG variables
 is in the opposite direction so the energy stays continuous and the path of the other RG variables is not affected.
\end{itemize}
In general, a trajectory of the RG variables contains all possible combinations of these events.
Some trajectories are very similar to the ones of the rational RG model with a positive interaction (see for
example the trajectory of the (011111100000) state Fig. \ref{poskop724}b,   
in contrast with negative interaction constant where this similarity is only present before the Moore-Read line.
For a nice example of recombinations see Fig. \ref{poskop724}c for the (010110101010) state. 
There is no condensate regime at positive interaction constant, and the RG variables don't need to become real for large $g$.
There is a category of trajectories which do not exist in the spectrum of the rational RG model
that we shall refer to as `suddencomplex' (sc) trajectories (see Fig. \ref{poskop724}d). In those trajectories we see that two real RG
variables suddenly become a complex conjugate pair with significant complex part, as opposed to the rational RG
model, where the formation of complex conjugate pairs of RG variables is a gradual process resulting from a singular point.
Finally we refer to Fig. \ref{poskop724}e for a nice combination of the different events described above.
Notice also that the energy stays continuous during all those trajectories as is required.
We found that the sc trajectories only occur above half filling, the RG variables under half filling remain real and analytical during the whole trajectory.
The fact that the Read-Green point at positive interaction constant only occurs
for filling fractions above half-filling could be related to this fact. The system at half-filling seems to have characteristics of a transitional region,
because the ground state has the same behaviour as below half-filling but some excited states start to
exhibit 'sudden collapses' and singular points as is typical for above half-filling (see Fig. \ref{poskop724}).
 \begin{figure*}[htb!]
 \includegraphics[width = 5cm]{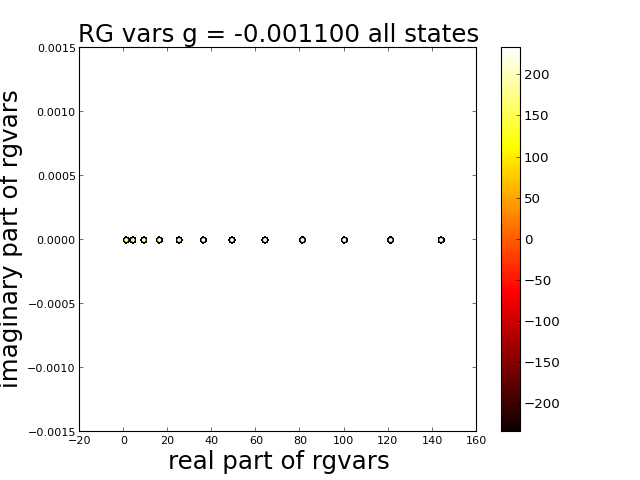}
 \includegraphics[width = 5cm]{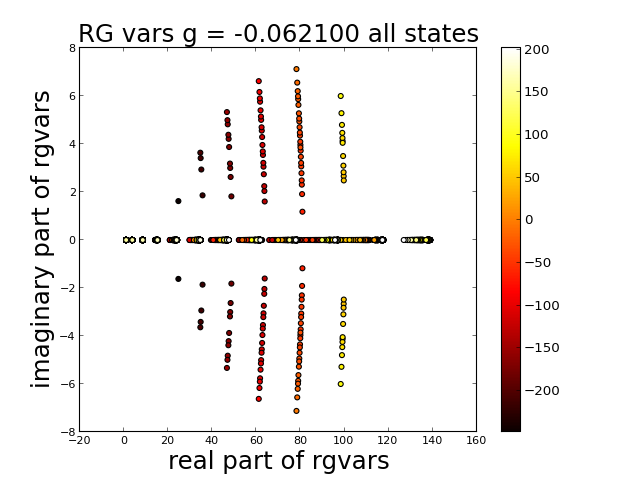}
 \includegraphics[width = 5cm]{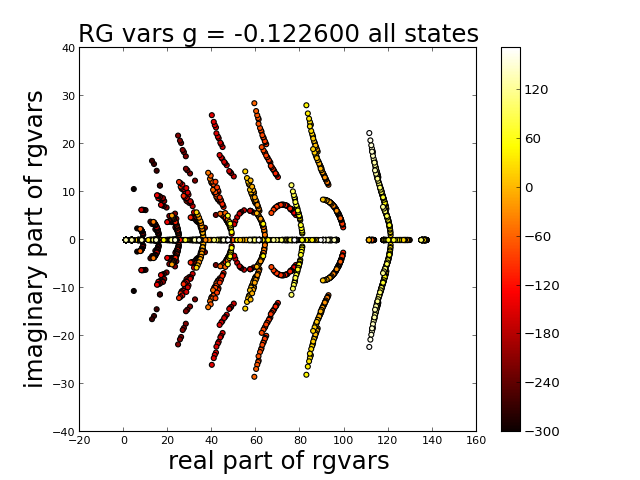}
 \includegraphics[width = 5cm]{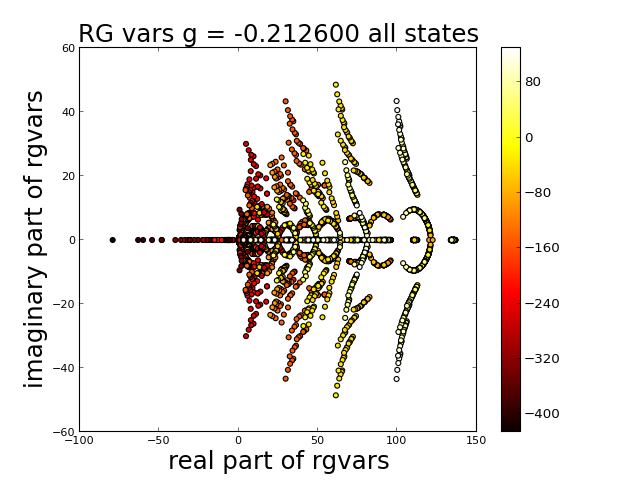}
 \includegraphics[width = 5cm]{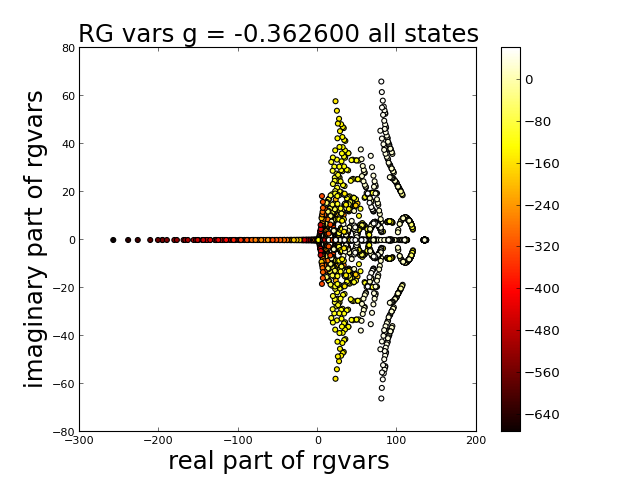}
 \includegraphics[width = 5cm]{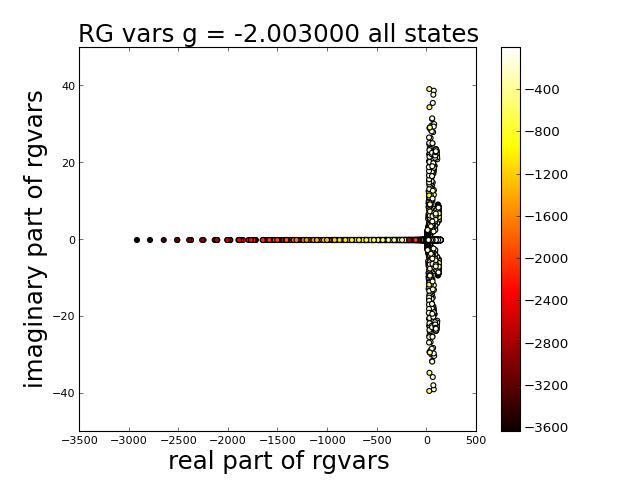}
 \caption{\label{fullspectrum} This figure shows all the RG variables of the full spectrum with 6 pairs in 12 doubly degenerate equidistant
 levels and zero seniority in function of increasing attraction strength. Colour coded according to the energy of the eigenstate
 to which they correspond.}
\end{figure*}

 \begin{figure}[htb!]
 \subfigure[$\quad$3 pair]{
  \includegraphics[width = 7.5cm]{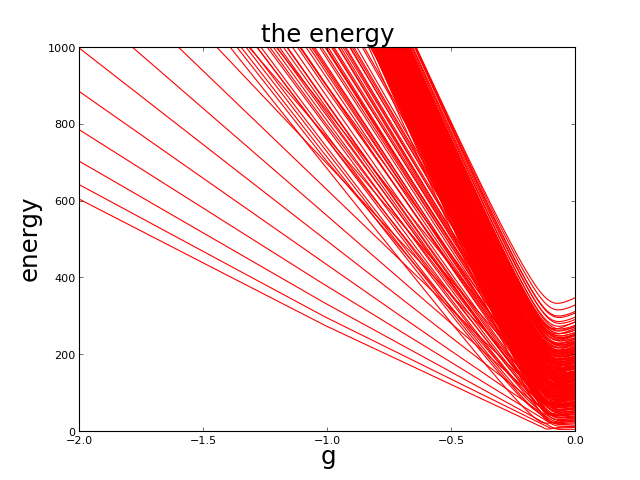}}
 \subfigure[$\quad$6 pair]{
  \includegraphics[width = 7.5cm]{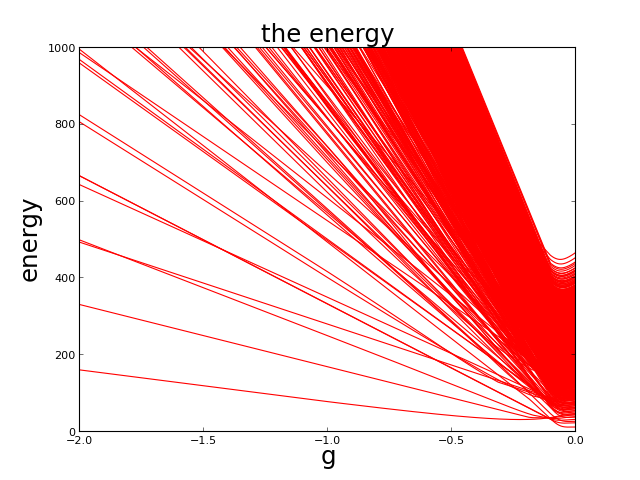}}
 \subfigure[$\quad$9 pair]{
 \includegraphics[width = 7.5cm]{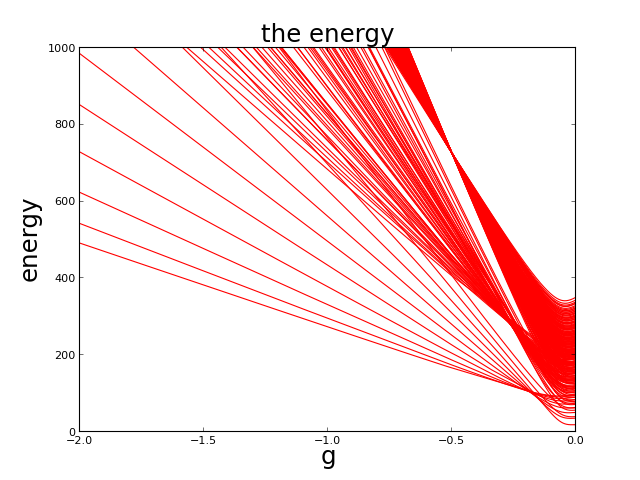}}
 \caption{All excitation energies of a system with 12 doubly degenerate single
 particle levels occupied by a) 3, b) 6 and c) 9 pairs and equidistant $D_i = i$ as a function of the interaction constant $g$.\label{nogap}}
\end{figure}

\begin{figure*}[H]
 \subfigure[$\quad$1 pair]{
  \includegraphics[width = 5.2cm]{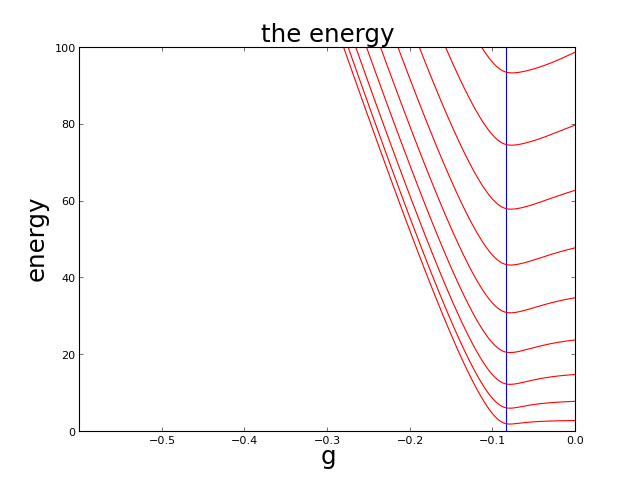}}
 \subfigure[$\quad$2 pair]{
  \includegraphics[width = 5.2cm]{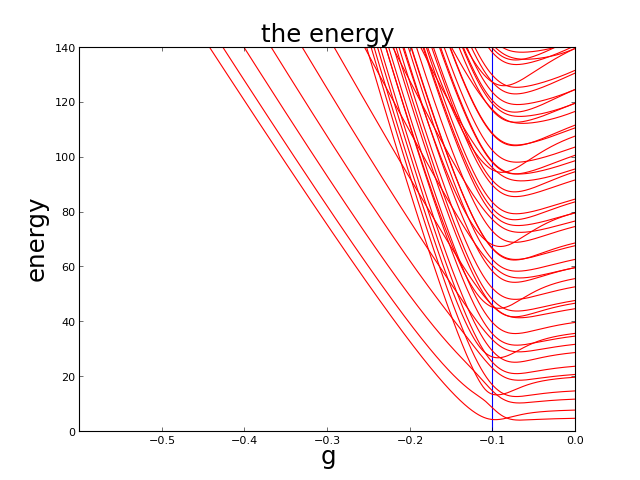}}
 \subfigure[$\quad$3 pair]{
 \includegraphics[width = 5.2cm]{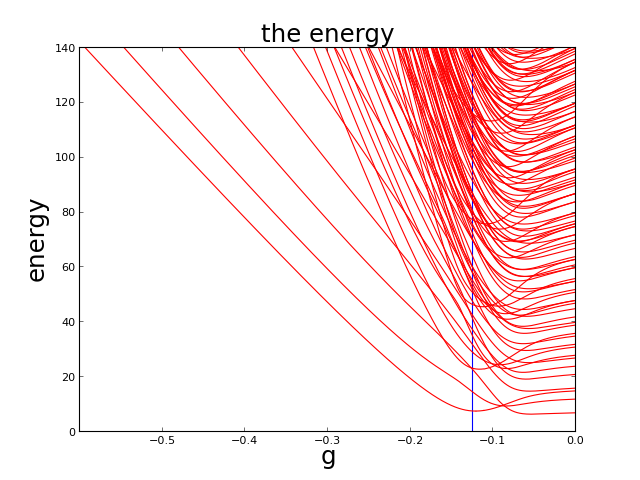}}
 \subfigure[$\quad$4 pair]{
 \includegraphics[width = 5.2cm]{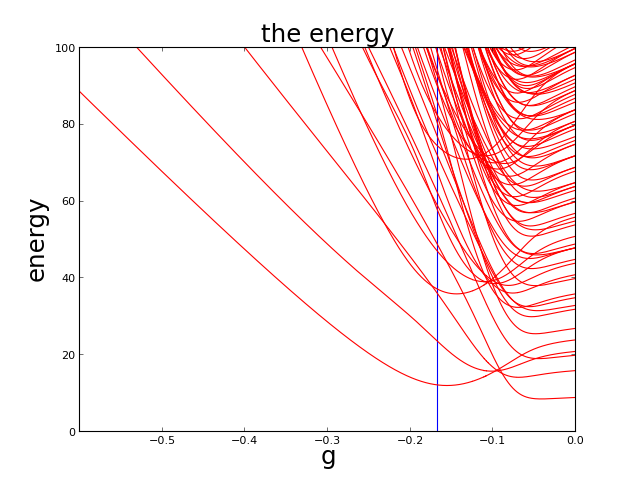}}
 \subfigure[$\quad$5 pair]{
 \includegraphics[width = 5.2cm]{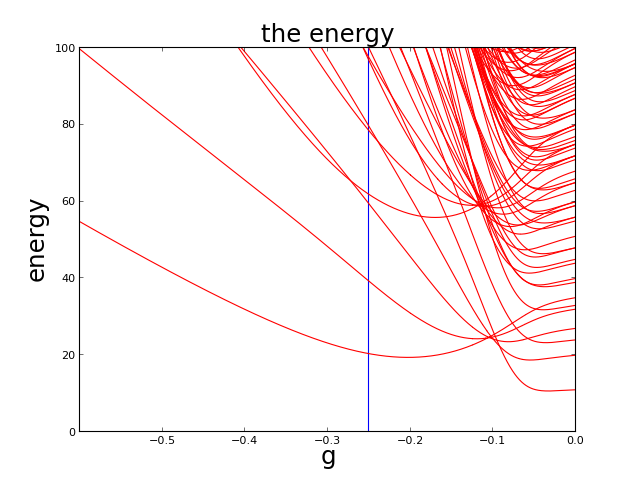}}
 \subfigure[$\quad$6 pair]{
 \includegraphics[width = 5.2cm]{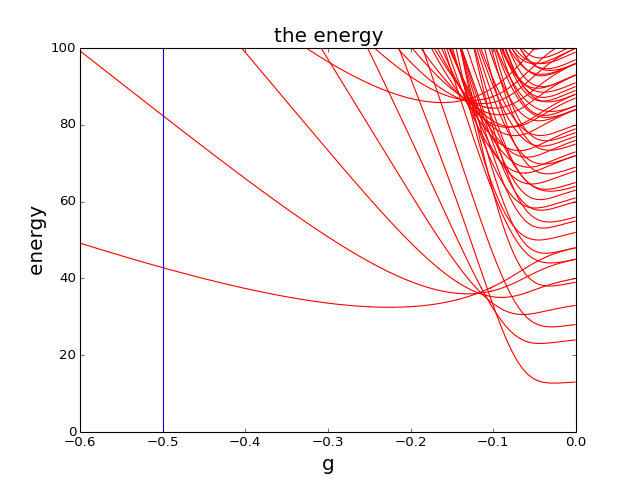}}
 \caption{The excitation energies of a system with 12 doubly degenerate sp levels
 occupied by 1 to 6 pairs (a-f) as a function of the interaction constant $g$. \label{readgreenfigure}
 The Read-Green point is depicted by a vertical line which divides the weak from the strong pairing regime.
 }
\end{figure*}
 
 \begin{figure}[htb!]
 \includegraphics[width = 8.5cm]{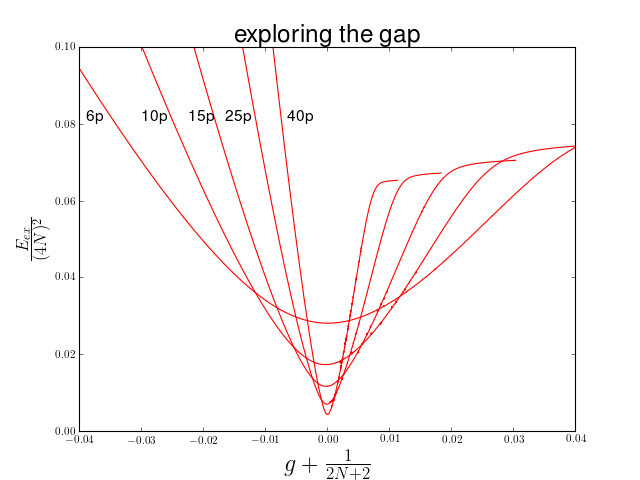}
 \caption{The energy differences between the ground state and the first excited state at the Read-Green point are depicted, this for several systems with an increasing number
 of pairs, all at quarter-filling, $D_i = i \; \text{with} \; i = 1 \dots 4N$ and $\eta =1$. The excitation energies are rescaled with a factor $(4N)^2$ and the interaction constant is shifted so the Read-Green point occurs for all systems at 0 (see eq. (\ref{conpoint})).  \label{gapsurvey}}
\end{figure}
 
\section{Excited states \label{exstates}}
\begin{table*}[htb!]
 \begin{tabular}[c]{|c|*{6}{l}|l}
  \hline 12 levels & $1 p$ & $2 p$ & $3 p$ & $4 p$ & $5p$ & $6p$  \\
  \hline \hline
   Read-Green point $g$ & -0.0833 & -0.1000& -0.1250& -0.1666 & -0.2500& -0.5000 \\
  TDA label & (010$\ldots$) &(0110$\ldots$) &(01110$\ldots$)&(011110$\ldots$) &(0111110$\ldots$)&(01111110$\ldots$) \\
  TDA Read-Green & (010$\ldots$) & (110$\ldots$)&(210$\ldots$) &(310$\ldots$) &(410$\ldots$) & (510$\ldots$) \\
  Energy  & 2.108110 & 4.323544& 7.257231& 11.627856& 20.383734& 42.779908 \\ 
   \hline 
 \end{tabular}
 \caption{\label{readgreentable}
The interaction constant ($g$) at the Read-Green point is calculated, for a system with 12 doubly degenerate sp levels and $\eta =1$. 
The first excited state reaches a minimum around the Read-Green point, the energy normalized to the ground-state energy at the Read-Green point is given, 
together with the start TDA distribution (label), and the TDA state that connects to the first excited state at the Read-Green point.}
\end{table*}

The above proposed algorithm to solve the RG equations is very robust and fast. This makes it possible to study entire spectra of mesoscopic systems.
It catches the eye that the RG variables of all states have the same typical evolution for changing interaction constant see Fig. \ref{fullspectrum}. The supplement of this paper contains
a movie that shows the evolution of the RG variables of all seniority zero states as the interaction strength is increased \cite{movie}.
Another interesting feature of the $p_x + ip_y$ Hamiltonian at half-filling is the fact that the gap between the ground state energy and the energy of the first excited state
is from the same order as gaps between higher excited states for an entire range of the interaction constant. Fig. \ref{nogap} depicts the entire
spectrum of a system with 12 doubly degenerate levels occupied by respectively 3,6 and 9 pairs with $\eta =1$. The Read-Green
line for 3 pairs in 12 levels is crossed at $g = -0.125$, which is exactly where the energy difference
of the ground state with the excited states reaches a local minimum and starts to increase rapidly.
Before the Read-Green line it is also possible for excited states to decrease the energy difference with the ground state, after the Read-Green line this
is not allowed any more.
This is also the case for 1, 2, 4, and 5 pairs as can be seen in Fig. \ref{readgreenfigure}. The TDA label associated
to the state with the local minimum excitation energy seems to have a pattern namely $(0\underbrace{1\ldots1}_{N}0\ldots)$, and the TDA state that connects to
the 1st excited state at the Read-Green interaction constant has the form $(N-1 \; 1 \;0 \ldots)$ (see Table \ref{readgreentable}). 
For 5 and 6 pairs, the minimum of the 1st excited state occurs a bit before the Read-Green point.
If the number of levels and pairs is increased while keeping the occupancy constant, the pattern remains and the increase of the excitation energies after the Read-Green point becomes much steeper, and
more and more states reach their minimum in excitation energy at the Read-Green point. 
In the continuum limit BCS theory predicts a strongly degenerate ground-state at the Read-Green point \cite{ibanez:2009}.
In order to investigate numerically the gap for growing system sizes approaching the thermodynamic limit, one would need to calculate a combinatorial number of excited states.
Because of the systematic $(0\underbrace{1\ldots1}_{N}0\ldots)$ TDA state labelling, observed in the small system (Fig. \ref{readgreenfigure}), we conjecture that the same
systematic holds for larger systems, so we only need to calculate two states to determine the minimum excitation energy at the Read-Green point.
Therefore it is possible to explore the behaviour of the gap 
for large system sizes. Fig. \ref{gapsurvey} shows the results for a system with an increasing number of pairs at quarter-filling. 
We take $D_i = i, \text{with} \; i = 1 \ldots 4N \; \text{for} \; N = 6 \ldots 40$. (Note that a full scan of the Hilbert space would require the calculation of $\binom{160}{40} \approx 8.6 \cdot10^{37}$ states for $N=40$.) The Read-Green point is predicted to be at $\frac{g}{\eta} = \frac{-1}{2N+2}$.
After rescaling the spectrum with $(4N)^2$, in order to guarantee a consistent definition of the thermodynamic limit with the 
highest sp level at $D_{4N}^2 = 1$, we see that the gap decreases for increasing system size as expected.
Another remarkable fact is that for bigger systems the gap, after the Read-Green point, increases much faster than for smaller systems.
This effect is stronger for lower filling fractions. 
Above half filling the system remains weakly paired over the entire range of the interaction constant and there is no hint of the formation of a gap. 

\section{Conclusions}
 In conclusion, we presented an efficient and stable method to solve a class of integrable pairing Hamiltonians. 
This makes it possible to probe entire spectra of systems with Hilbertspaces way beyond the realm of exact diagonalisation
 techniques. The method solves the Bethe ansatz equations by means of a deformation parameter which adiabatically
 connects the genuine boson limit to the hard-core boson limit. Furthermore, we related the singular points of the RG variables to a change in the associated TDA distribution and corresponding overlaps. The ground state connects with the most collective TDA state slightly before the Moore-Read line. 
 In the low interaction regime, the path of the RG variables of the factorisable interaction has some resemblances with the 
 reduced BCS Hamiltonian which also appeared in the overlaps with the bosonic states. However, after the low interaction regime, an entirely different regime arises which has
no resemblance with a regime of the reduced BCS case. Remnants of the Read-Green line for finite size systems are found as a local minimum of the
first excited state, before the Read-Green point excited states can lower their energy difference with the ground-state energy after the 
Read-Green point this is no longer possible. Finite-size effects cause this minimum to shift to weaker interaction strength when half-filling is approached, with an
increasing amount of sp levels this shift gets noticeable for higher filling fractions only.
A pattern is found for the label of the TDA state that becomes the first excited state at
the Read-Green point and the TDA state that connects to the first excited state at the Read-Green point. 
Future investigations could look for a pattern in the TDA start distributions which will lead to hard-core boson solutions
of the factorisable Hamiltonian, and which TDA start distributions are not linked with hard-core boson solutions at a particular 
interaction constant and at which $\xi$ value they break down, or extending the above approach to other integrable models.

\begin{acknowledgments}
SDB is a `FWO-Vlaanderen' postdoctoral
 fellow. Patrick Bultinck is acknowledged for continued support during the final stages of this work.
\end{acknowledgments}

\appendix
\section{The near-contraction limit \label{nclim}}
In this appendix we derive an approximate solution to the generalised Richardson-Gaudin equations eq.(\ref{rgeqdef}) for very small $\xi$.
Recall that the RG equations with $\xi = 0$ are given by:
\begin{equation}
1 + \frac{g}{2} \sum_i \frac{ D_i^2 \Omega_i}{\eta D_i^2  - E_{\alpha}(0)} = 0 \label{xi0}.
\end{equation}
The following form of the RG variables for very small $\xi$ is assumed. 
\begin{equation}
 E_{\alpha}\left(\xi\right) = E_{\alpha}\left(0\right) + \sqrt{\xi} x_{\alpha}
\end{equation}
$\xi$ is chosen very small so it is possible to perform a series expansion in $\sqrt{\xi}$ in the second term of eq.(\ref{rgeqdef})
\begin{multline}
 1+2g \sum_i \frac{D_i^2 \left(\frac{1}{4} \Omega_i - \frac{1}{2} \xi v_i \right)}{\eta D_i^2 - E_{\alpha}\left(0\right)} \\ 
 \biggl[1+ \frac{\sqrt{\xi}x_{\alpha}}{\eta D_i^2 - E_{\alpha}\left(0\right)} + \ldots \biggr] \\
- 2 \frac{\xi g}{\eta} \sum_{\beta \neq  \alpha} \frac{E_{\beta}\left(0\right)+  \sqrt{\xi}x_{\beta}}{E_{\beta}\left(0\right) - E_{\alpha}\left(0\right) + \sqrt{\xi} \left(x_{\beta} - x_{\alpha}\right)} \approx 0.
 \end{multline}
 Now we split the summation of the third term in the above equation into a part for which $E_{\beta}\left(0\right) = E_{\alpha}\left(0\right)$ and
 a part for which $E_{\beta'}\left(0\right) \neq E_{\alpha}\left(0\right)$.
\begin{multline}
 1+2g \sum_i \frac{ D_i^2 \left(\frac{1}{4} \Omega_i - \frac{1}{2} \xi v_i \right)}{\eta D_i^2 - E_{\alpha}\left(0\right)} \\
 \biggl[1+ \frac{\sqrt{\xi}x_{\alpha}}{\eta D_i^2 - E_{\alpha}\left(0\right)} + \ldots \biggr] \\
- 2 \frac{\sqrt{\xi} g}{\eta} \sum_{\beta \neq \ \alpha} \frac{E_{\alpha}\left(0\right)+\sqrt{\xi}x_{\beta}}{ \left(x_{\beta} - x_{\alpha}\right)} \\
- \frac{2\xi g}{\eta} \sum_{\beta' \neq \ \alpha} \frac{E_{\beta'}\left(0\right) 
+\sqrt{\xi}x_{\beta'}}{E_{\beta'}\left(0\right) - E_{\alpha}\left(0\right) + \sqrt{\xi} \left(x_{\beta'} - x_{\alpha}\right)}\approx 0.
\end{multline}
After gathering the terms of order $O\left(1\right)$, we see that they are zero 
because of eq.(\ref{xi0}). For the $O\left(\sqrt{\xi}\right)$ terms we obtain:
\begin{equation}
 ax_{\alpha} + \frac{E_{\alpha}\left(0\right)}{\eta} \sum_{\beta \neq \alpha} \frac{2}{x_{\alpha} -  x_{\beta}} = 0 \label{stieltjeap}
\end{equation}
The index $\beta$ runs only over the $n$ indices such that $E_{\beta}\left(0\right)= E_{\alpha}\left(0\right)$, and $a = \frac{1}{2}\sum_i \frac{ D_i^2 \Omega_i }{\left(\eta D_i^2 - E_{\alpha}\left(0\right) \right)^2}$.
 The equation above is of the Stieltjes type \cite{stieltje}, so we can define a Stieltjes polynomial.
 \begin{equation}
  P\left(x\right) = \prod_{i}^{n}\left(x-x_{i}\right),
 \end{equation}
with $x_{i}$ the roots of the Stieltjes equations. Remark that $\sum_{j \neq i}^n \frac{2}{x_{i} - x_{j}} = \frac{P''(x)}{P'(x)}$,
multiply eq.(\ref{stieltjeap}) with $P'(x)$ and take into account the fact that polynomials of the same order, with the same zeros are equal
up to a scale factor which in this case is $a n$. 
This gives finally the following corresponding differential equation.
\begin{equation}
 \frac{E_{\alpha}\left(0\right)}{\eta} P''\left(x\right) + axP'\left(x\right) = a n P\left(x\right) 
\end{equation}
If we now apply the transformation $z = i \sqrt{\frac{\eta a}{2E\left(0\right)}}x$,
we can transform this equation into a `physicists' Hermite differential equation.
\begin{equation}
 H'' - 2zH'\left(z\right) + 2 n H\left(z\right) = 0
\end{equation}
So finally we get for the $E_{\alpha}\left(\xi\right)$ variables in the $\xi \rightarrow 0$ limit:
\begin{equation}
 E_{\alpha}\left(\xi\right) \approx E_{\alpha}\left(0\right) + i \sqrt{\frac{2E_{\alpha}\left(0\right)\xi}{\eta a}} z_{\alpha}^{\nu} \quad \xi \ll 1,\quad \nu = 1\ldots n \label{nearTDA}
\end{equation}
with $z_{\alpha}^{\nu}$ the $\nu$-th root of the `physicists' Hermite polynomial $H_{n}\left(z\right)$.

\section{Condensation points \label{conap}}
We determine the condition for which $p$ pairs with zero energy and $q$ general RG pairs
form an eigenstate of the  $p_x + ip_y $ pairing Hamiltonian (see eq.(\ref{hfac})).
For the $E_{\alpha} = 0$ pairs the generalised pair operators become:
\begin{eqnarray}
 \da{K}_0 &=& \sum_{k = 1}^m \frac{\da{S}_k}{\eta D_k^*}, \quad K_0 = \left(\da{K}_0 \da{\right)} \\
 K_0^0 &= &\sum_{k = 1}^m \frac{S_k^0}{\eta}
\end{eqnarray}
So we have to derive under which conditions the state:
\begin{equation}
 \ket{\psi} = \left(\da{K}_0 \right)^p \prod_{\alpha = 1}^q \da{K}_{\alpha} \ket{\theta} \label{constate}
\end{equation}
is an eigenstate of the factorisable Hamiltonian (\ref{hfac}). This will be done by commuting the Hamiltonian through the product state (\ref{constate}),
and break down the resulting state into the eigenstate and the orthogonal part. 
Pulling the Hamiltonian through the $p$ condensed pairs gives:
\begin{multline}
 H \left(\da{K}_{0} \right)^p
 = \frac{1}{2} p \left(p-1\right) \left( \da{K}_0 \right)^{p-2}  \left[ \left[ H , \da{K}_0 \right], \da{K}_0 \right] \\ + p \left( \da{K_0} \right)^{p-1} \left[ H , \da{K}_0 \right] + \left( \da{K}_0 \right)^p H
 \end{multline}
The commutators in the above expression are given by:
\begin{eqnarray}
 \left[ H , \da{K}_0 \right] &=& -2 \frac{g}{\eta} \da{K}_D \da{K}_0 \\
 \left[ \left[ H , \da{K}_0 \right], \da{K}_0 \right] &=& \da{K}_D \left(1 - 2 \frac{g}{\eta} K_0^0 \right) 
\end{eqnarray}
Where $\da{K}_D = \sum_k D_k \da{S}_k$.
We already know how the Hamiltonian commutes through the product state $\prod_{\alpha = 1}^q \da{K}_{\alpha}$ yielding the RG equations for $q$ pairs, so we only need to calculate the additional 
commutator.
\begin{equation}
 \left[K_0^0 , \prod_{\alpha = 1}^q \da{K}_{\alpha} \right]= q \prod_{\alpha =1 }^q \da{K}_{\alpha}
\end{equation}
At the end we get the following relation;
\begin{multline}
 H \left( \da{K}_0 \right)^p \left( \prod_{\alpha = 1}^q \da{K}_{\alpha} \right) \ket{\theta} = \left( \da{K}_0 \right)^p H \left( \prod_{\alpha = 1}^q \da{K}_{\alpha} \right)\ket{\theta}  \\ 
 + \left(\da{K}_0\right)^{p-1} \da{K}_D \prod_{\alpha =1}^q \da{K}_{\alpha} \\ \left[ -2 \frac{g}{\eta}p q - \frac{g}{\eta} p \left( p-1 \right) + p \left( 1 + 2 \frac{g}{\eta} \sum_{k=1}^{m} s_k \right) \right] \ket{\theta}
\end{multline}
The first line corresponds to the standard RG equations for the $q$ remaining pairs, whereas the second line gives an additional constraint if we 
want the state eq.(\ref{constate}) to be an eigenstate:
\begin{equation}
 \frac{\eta}{g} = 2q + p -1 -2 \sum_{k =1}^m s_k. \label{conp}
\end{equation}
RG variables are only allowed to 'condense' at $N$ discrete ratios of the interaction constant $g$ and $\eta$ if one of the two is held constant, where $N$ is the total number of pairs present in the system under investigation,
because the number $p$ of condensed pairs can be any number between zero and $N$ and $q = N-p$.

\section{Around the condensation points \label{rgcon}}
At the condensation points (\ref{conp}), $p$ of the $N$ RG variables are condensed to zero, leading to singularities in the RG equations (\ref{rgeq}).   However, it is possible to extract the qualitative behaviour of the RG variables around the condensation points by expanding the RG equations (\ref{rgeq}) around the condensation points (\ref{conp}).  For our purpose, it is convenient to rewrite the RG equations (\ref{rgeq}) in the following form
\begin{align}
\frac{1}{E_\alpha}&\left[\frac{\eta}{2g}+\sum_{i}s_i-(N-1)\right]+\sum_{i}\frac{s_i}{\eta D_i^2-E_\alpha}\notag\\
&-\sum_{\beta\neq\alpha}\frac{1}{E_\beta-E_\alpha}=0,\qquad\forall \alpha=1\dots N.
\end{align}
Expanding the interaction constant $g+\delta g$ around the condensation points eq.(\ref{conp})
\begin{equation}
\frac{\eta}{g+\delta g}\approx\frac{\eta}{g}-\frac{\eta\delta g}{g^2}
\end{equation}
the equations become
\begin{align}
\frac{1}{E_\alpha}&\left[\frac{-p+1}{2}-\frac{\eta\delta g}{g^2}\right]+\sum_{i}\frac{s_i}{\eta D_i^2-E_\alpha}\notag\\
&-\sum_{\beta\neq\alpha}\frac{1}{E_\beta-E_\alpha}=0,\qquad\forall \alpha=1\dots N.
\end{align}
It is reasonable to assume that the $p$ condensed variables $E_\alpha$ (with $\alpha=1\dots p$) in the vicinity of the condensation points can be developed in a series expansion of $\delta g^\gamma$ with $\gamma$ a yet unknown exponent, whereas the other $q=N-p$ variables can be assumed finite.
\begin{equation}
E_\alpha=\left\{\begin{array}{ll}
   x_\alpha \delta g^\gamma & \alpha=1\dots p\\
   y_\alpha & \alpha=p+1\dots N\end{array}\right.
\end{equation}
The equations break down into two coupled sets 
\begin{align}
&\frac{1}{x_\alpha \delta g^\gamma}\left[\frac{-p+1}{2}-\frac{\eta\delta g}{g^2}\right]+\sum_{i}\frac{s_i}{\eta D_i^2-x_\alpha \delta g^\gamma}\notag\\
&\quad-\sum_{\beta\neq\alpha}^p\frac{1}{(x_\beta-x_\alpha)\delta g^\gamma}-\sum_{\beta=p+1}^N\frac{1}{y_\beta -x_\alpha\delta g^\gamma }=0\label{rgeq:condensationpoints},\\
&\frac{1}{y_\alpha }\left[\frac{-p+1}{2}-\frac{\eta\delta g}{g^2}\right]+\sum_{i}\frac{s_i}{\eta D_i^2-y_\alpha }\notag\\
&\quad-\sum_{\beta=1}^p\frac{1}{x_\beta\delta g^\gamma-y_\alpha}-\sum_{\beta=p+1\neq\alpha}^N\frac{1}{y_\beta -y_\alpha}=0\label{rgeq:noncondensationpoints},
\end{align}
with the first set (\ref{rgeq:condensationpoints}) related to the condensed variables ($\alpha=1\dots p$) and the second set (\ref{rgeq:noncondensationpoints}) referring to the non-condensed variables  ($\alpha=p+1\dots N$).  In lowest order in $\delta g^\gamma$, these equations become decoupled
\begin{align}
&\frac{p-1}{2x_\alpha}+\sum_{\beta\neq\alpha}^p\frac{1}{x_\beta-x_\alpha}=0,\quad\forall \alpha=1\dots p\label{rgeq:condensationpoints:zerothorder}\\
&\frac{p+1}{2y_\alpha }+\sum_{i}\frac{s_i}{\eta D_i^2-y_\alpha }- \sum_{\beta=p+1\neq\alpha}^N\frac{1}{y_\beta -y_\alpha}=0, 
\end{align}
The latter set of equations depend on the parameters in the model, whereas the former set is purely geometric.  It can be shown that the variables $x_\alpha$ are located at the corners of a regular $p$-polygon in the complex plain.
\begin{equation}\label{rqeq:condensationpoints:polygonsolution}
x_\alpha = x_0 \omega^{\alpha-1},\qquad\forall\alpha=1\dots p
\end{equation}
with $\omega^p=1$.  Substituting (\ref{rqeq:condensationpoints:polygonsolution}) into  (\ref{rgeq:condensationpoints:zerothorder}) yields the set of equations
\begin{equation}
\frac{p-1}{2}+\sum_{\beta\neq\alpha}^p\frac{1}{\omega^{\beta-\alpha}-1}=0,\quad\forall \alpha=1\dots p
\end{equation}
Because of the periodicity $\omega^{\alpha+p}=\omega^{\alpha}$, this set of equations is equivalent to one single equation
\begin{equation}
\frac{p-1}{2}+\sum_{\beta=1}^{p-1}\frac{1}{\omega^{\beta}-1}=0,
\end{equation}
which can be shown to hold identically for periodic solutions $\omega^p=1$.  As a result, the variables $x_\alpha$ around the condensation point approach $x_\alpha=0$ along the corners of a regular $p$-polygon (See Figure \ref{mrclose}).  It is worth pointing out that the geometric solution (\ref{rqeq:condensationpoints:polygonsolution}) is independent of the free variable $x_0$ or the scaling parameter $\gamma$, for which higher orders in the series expansion should be considered.  We leave this for further investigations.
\bibliographystyle{chicago}
\bibliography{mybib}
\end{document}